\documentclass[10pt,prl,onecolumn,superscriptaddress]{revtex4}

\usepackage{graphicx}
\usepackage{amssymb}
\usepackage{times}
\usepackage{amsmath}
\usepackage{amsfonts}
\usepackage{txfonts}
\usepackage{color}

\begin{document}

\title{Back-scatter based  whispering gallery mode sensing}

\author{Joachim Knittel} \affiliation{School of Mathematics and Physics, University of Queensland, St Lucia, QLD 4072, Australia} 
\author{Jon D. Swaim} \affiliation{School of Mathematics and Physics, University of Queensland, St Lucia, QLD 4072, Australia} %

\author{David L. McAuslan} \affiliation{Centre for Engineered Quantum Systems, School of Mathematics and Physics, University of Queensland, St Lucia, Brisbane, QLD 4072, Australia}

\author{George A. Brawley} \affiliation{Centre for Engineered Quantum Systems, School of Mathematics and Physics, University of Queensland, St Lucia, Brisbane, QLD 4072, Australia}

\author{Warwick P. Bowen} \affiliation{Centre for Engineered Quantum Systems, School of Mathematics and Physics, University of Queensland, St Lucia, Brisbane, QLD 4072, Australia}

\begin{abstract}
Whispering gallery mode biosensors allow selective unlabelled detection of single proteins and, combined with quantum limited sensitivity, the possibility for noninvasive real-time observation of motor molecule motion. However, to date technical noise sources, most particularly low frequency laser noise, have constrained such applications. Here we introduce a new technique for whispering gallery mode sensing based on direct detection of back-scattered light. This experimentally straightforward technique is immune to frequency noise in principle, and further, acts to suppress thermorefractive noise. We demonstrate 27~dB of frequency noise suppression, eliminating frequency noise as a source of sensitivity degradation and allowing an absolute frequency shift sensitivity of 76~kHz. Our results open a new pathway towards single molecule biophysics experiments and ultrasensitive biosensors.
\end{abstract}

\date{\today} \maketitle

\begin{figure*}[b!]
\begin{center}
\includegraphics[width=\textwidth]{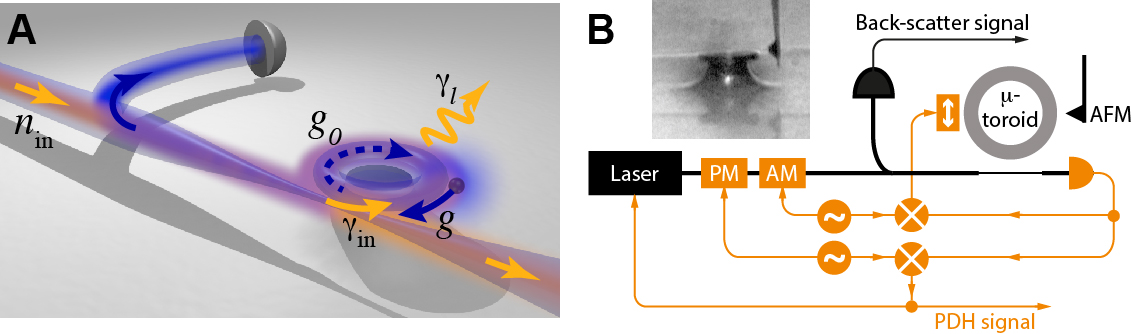}
\caption{{\bf Back-scatter based WGM biosensor.} {\bf A}: Illustration of concept. Resonator loss rate: $\gamma_l = \gamma - \gamma_{\rm in}$. {\bf B}: Schematic  of experiment. Black components: critical components for back-scatter measurement. Orange components: components used for locking and direct frequency measurement. Inset: optical micrograph of sensor. PM: Phase modulator. AM: Amplitude modulator. Orange box containing vertical double-sided arrow: nano-positioning stage used to control separation of taper and toroid.
}
\label{schematic}
\end{center}
\end{figure*}

Whispering gallery mode (WGM) biosensors combine high quality optical resonators with surface functionalisation to allow ultra-sensitive detection of unlabelled biomolecules such as proteins and nucleic acids\cite{Vollmer-2002, Vollmer-2003}. When a biomolecule enters the evanescent field of the sensor, it becomes polarised and the sensor's optical resonance experiences a reactive shift to lower frequency\cite{Arnold-2003}. This frequency shift can be directly measured with high precision using the optical field exiting the resonator. Operation at the fundamental noise limit dictated by the combination of quantum shot noise and thermo-refractive noise could in principle enable single binding events of unlabelled proteins to be easily resolved, and real-time observation of conformational changes and motor molecule motion\cite{Topolancik-2006}. These capacities have significance for many applications in fundamental science, industry, and medical diagnosis. However, such applications have not been possible to date, with technical noise constraining all WGM biosensors to sensitivities many orders of magnitude worse than the fundamental limit. Several avenues have been developed to increase the single-molecule frequency shift and thereby elude technical noise sources, including most notably the use of electric field hot-spots created by plasmonic nanoparticles deposited on the resonator\cite{Shopova-2011, Santiago-Cordoba-2011, Swaim-2011, Dantham-2013} and of thermo-optic heating of the resonator by the biomolecule itself\cite{Armani-2007}. In both of these cases, single molecule sensitivity has been achieved\cite{Dantham-2013}.  However, similarly to more conventional reactive sensing, the sensitivity of these experiments is dominated by laser frequency noise and remains orders of magnitude above the fundamental limit.

Over the past few years significant efforts have been made to reduce technical noise sources, and most particularly laser frequency noise. In a notable example, direct laser frequency noise cancellation has been achieved using an independent reference interferometer to characterise the laser frequency noise in real-time as experiments proceed\cite{Lu-2011}.  This allowed the detection of 12.5~nm radius nanoparticles with a frequency shift noise floor of 133~kHz. Self-referencing provides an elegant alternative approach to suppress frequency noise, and is particularly well suited to WGM sensors since they inherently exhibit frequency degenerate pairs of forwards- and backwards-propagating optical modes. The presence of a nanoparticle or biomolecule introduces scattering between these modes, breaking the degeneracy and causing a frequency split pair of orthogonal standing wave modes to form in the resonator\cite{He-2011}. Measuring the splitting then provides a self-referenced method to sense the presence of the scatterer that is in principle frequency noise immune. This approach was first demonstrated in Ref.~\cite{Zhu-2010}, where nanoparticles of 30 nm radius were detected in air by scanning laser spectroscopy of the frequency splitting. However, such spectroscopic techniques are constrained to frequency splittings larger than the resonator optical decay rate, where the splitting becomes spectroscopically resolvable. In the best WGM biosensing experiments to date, the optical decay rate has been in the range of a few MHz\cite{Armani-2007} precluding the possibility of detecting single proteins or other small molecules. 
Proof-of-principle experiments in air have recently overcome this constraint both by embedding the resonator within a Sagnac interferometer and thereby allowing independent measurements of the frequency of each standing wave mode\cite{Knittel-2010}; and by doping the resonator material with a laser medium, pumping it such that both modes exhibit lasing, and directly measuring the beat frequency between them\cite{He-2011}. To the authors knowledge, the latter of these experiments achieved the lowest frequency noise floor previously reported in any WGM sensor of 100~kHz. However, both of these approaches involve a substantial increase in complexity over conventional methods which restricts practical applications. 

Here, we report a new approach to self-referenced real-time WGM biosensing that can be straightforwardly applied within conventional sensing configurations. The essence of the approach is to directly monitor the intensity of back-scattered light from the resonator. By scattering light from the forwards- to the backwards-propagating resonator modes, the presence of a biomolecule or nanoparticle directly affects this intensity. We show theoretically that this approach is immune to laser frequency noise in principle; and  that, when compared to a direct frequency shift measurements on a single cavity mode, achieves the same quantum shot noise sensitivity limit, with thermorefractive noise suppressed by a factor of two. A proof-of-principle experiment is implemented using an atomic force microscopy (AFM) tip to controllably mimic the presence of a biomolecule\cite{Knittel-2013}. Twenty seven~dB of frequency noise suppression is achieved, eliminating frequency noise as a limiting factor for the sensitivity of the sensor.  This compares to 20~dB reported in Ref.~\cite{Lu-2011}. In that work, however, a large external interferometer was required which was thermally and mechanically insulated  and stabilised for several hours in an ice-water bath. By contrast, in our work, no additional optical or electronic systems are required. The absolute frequency noise floor achieved in our experiments is 76~kHz, which to our knowledge is superior to all previous WGM sensors\cite{He-2011}. Furthermore, the approach is directly compatible with enhanced WGM sensing approaches using both plasmonic field hot-spots\cite{Shopova-2011, Santiago-Cordoba-2011, Swaim-2011, Dantham-2013} and thermo-optic heating\cite{Armani-2007}.

As shown in Fig.~\ref{schematic}A, we consider a WGM resonator with total decay rate $\gamma$, coupled to an external optical field at rate $\gamma_{\rm in}$. The optical field is detuned a frequency $\Delta$ from the optical resonance frequency. An intrinsic scattering rate $g_0$ is included between the forwards- and backwards-propagating whispering gallery modes, which in experiments typically results from surface roughness and Rayleigh scattering. The backscattered field arise both from this intrinsic scatter and from scattering introduced by the possible presence of a biomolecule or nanoparticle. The frequency shift experienced by the standing-wave mode that interacts with the biomolecule or nanoparticle is exactly equal to twice the scattering rate $g$ it introduces between counter-propagating modes\cite{Zhu-2010}. Consequently, a measurement of the scattered intensity directly yields information about the the frequency separation of the split standing wave modes and thereby the polarisability of the scatterer. In the Supplementary Information we calculate the frequency noise spectrum $S(\omega)$ of a measurement of $g$ in the realistic limit that $\{ \Delta, g_0 \} \ll \gamma$ and $g \ll g_0$, with the result being
\begin{equation}
S(\omega) = \left ( \frac{g_0}{2} \right )^2 S_{\rm RIN} (\omega)  +  \left ( \frac{g_0 \Delta}{\gamma}\right )^2 \! \! S_{\omega} (\omega)  +  \frac{1}{2} S_{\rm T} (\omega) + S_{\! \rm shot}, \label{eqn}
\end{equation}
where $S_{\! \rm shot} =  \gamma^4/16 n_{\rm in} \gamma_{\rm in}^2$ is the quantum shot noise floor of the measurement, with $n_{\rm in}$ being the incident photon flux; $S_T(\omega)$ is the thermorefractive noise floor for measurements on one circulating WGM mode;
$S_{\rm RIN} (\omega)$ is the power spectrum of relative intensity noise of the incident laser;
and $S_{\omega} (\omega)$ is the power spectrum of the laser frequency noise.
As can be seen, for zero detuning ($\Delta=0$) laser frequency noise is perfectly suppressed. This is in distinct contrast to conventional approaches to WGM biosensing that rely on direct measurements of resonant frequency shifts. In that case, laser frequency noise is indistinguishable from signals due to the presence of the biomolecule or nanoparticle (see Supplementary Information). 

When $\Delta=0$ the remaining noise in Eq.~(\ref{eqn}) above is due only to quantum shot noise, laser relative intensity noise, and thermorefractive noise. Somewhat remarkably, the quantum shot noise contribution is exactly identical to that of an ideal conventional measurement, while the thermorefractive noise is suppressed by a factor of two (see Supplementary Information for an explicit comparison). This thermorefractive noise suppression can be understood since in the back-scatter configuration the thermorefractive noise arises from back-scattering due to thermally driven local variations in refractive index. This causes a combination of both amplitude and phase fluctuations in the total back-scattered field, with the ratio dependent on the spatial origin of the thermorefractive back-scattering. Averaged over all possible origins within the WGM resonator, the fluctuations are imprinted equally on amplitude and phase. However, only the amplitude fluctuations contribute to the noise floor. The Supplementary information also considers noise due to input coupling fluctuations, arising for example from variations in the distance between the resonator and the coupling device. To first order, this noise is perfectly suppressed in the usual operating regime of critical coupling, where the incident optical field is fully coupled into the resonator. 

To test the predictions of the model, we implemented the experiment shown in Fig.~\ref{schematic}B. The WGM biosensor consisted of a silicon chip based microtoroidal resonator with optical quality factor of approximately $2\times10^7$, optically excited with 4.3~$\mu$W of 1557.5~nm laser light via a tapered optical fibre. A Thorlabs 6015-3-APC circulator placed in the fibre line prior to the resonator allowed the backscattered intensity to be measured on a New Focus 2011 photoreceiver, with approximately 50~nW of backscattered power observed when the laser was on resonance.  To compare the effectiveness of this backscatter measurement to conventional frequency shift measurements, the transmitted intensity was also detected which, combined with a laser phase modulation, allowed a Pound-Drever-Hall (PDH) type error signal to be generated quantifying the frequency shift\cite{Swaim-2013}. This error signal additionally allowed the laser, a New Focus Velocity external cavity diode laser, to be locked onto the optical resonance frequency via feedback to its internal PZT, enforcing the zero-detuning condition $\Delta=0$. 
To further stabilise the experiments, the optical coupling rate to the resonator was locked to critical coupling by amplitude modulating the laser to extract a zero-crossing error signal, and feeding this back to a PZT controlling the position of the tapered optical fibre\cite{Chow-2012}. 

\begin{figure}
\begin{center}
\includegraphics[width=0.47\textwidth]{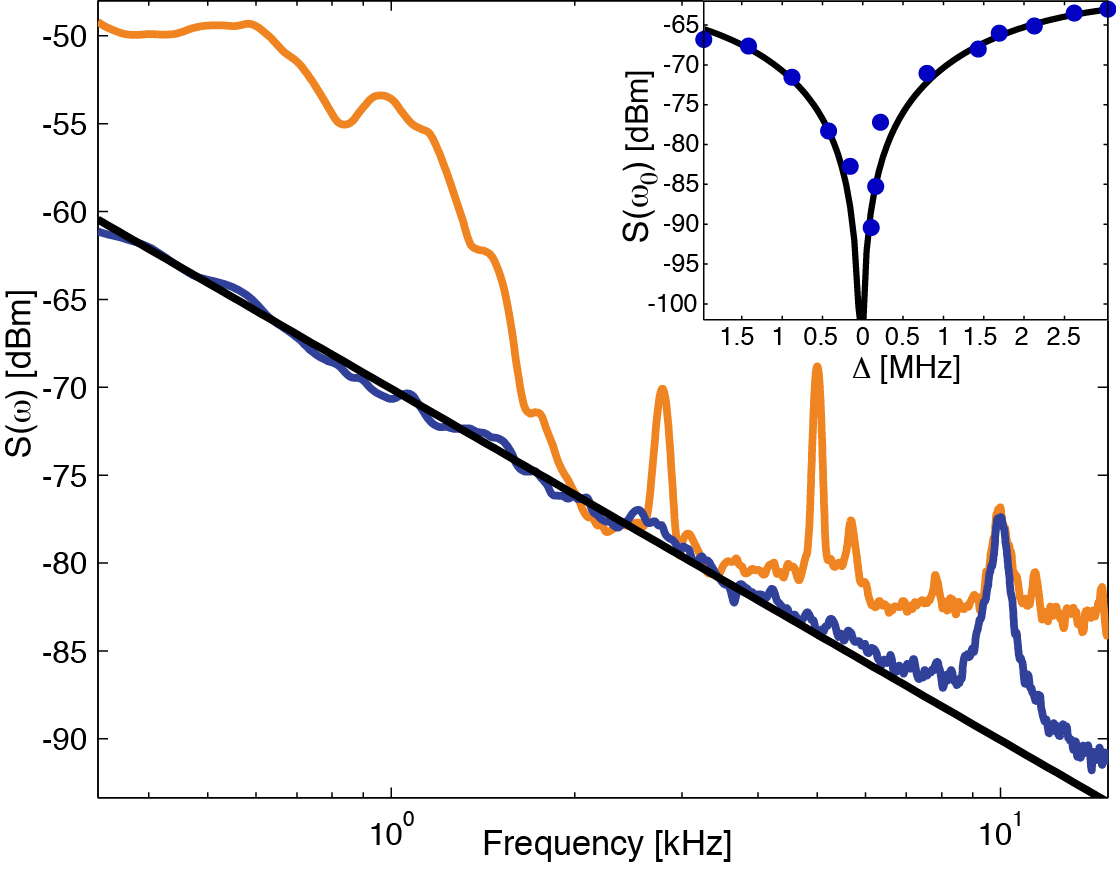}
\caption{{\bf Frequency noise comparison.} Frequency noise spectra for back-scatter based (blue/dark trace) and PDH based (orange/light) measurements. Black line: $1/f^2$ fit to back-scatter data. Inset: Back-scattered power spectrum at modulation frequency as a function of detuning. Black curve: fit to theory in supplementary information. $\omega_m$: 16~kHz modulation frequency.}
\label{FreqNoise}
\end{center}
\end{figure}
In an initial experiment, the relative frequency noise spectra for backscatter and direct frequency shift measurement were obtained using spectral analysis of the back-scattered intensity and the PDH error-signal. The presence of a nanoparticle or biomolecule was controllably simulated by using a nano-positioning stage to place a standard AFM tip within the evanescent field of the toroid. A  modulation applied to the position of the AFM tip at it's fundamental resonance frequency of 9.7~kHz caused a sinusoidal variation in both backscattering rate and frequency shift. By equating the power spectra from the two measurements at this frequency, the relative noise floors of the two approaches could be calibrated.  The resulting calibrated spectra are shown in Fig.~\ref{FreqNoise}. As can be seen, the noise floor of the back-scatter measurement is lower than direct measurement over the full frequency window, and particularly at the low frequencies relevant to biosensing experiments where laser frequency noise becomes increasingly significant. The Lorentzian peak evident at 9.7~kHz in both spectra is the AFMs thermally excited fundamental mechanical mode. With the exception of this peak, the back-scatter measurement is dominated by $1/f^2$ noise with no other technical noise evident. We believe that this noise is laser intensity noise, with separate measurements of the laser source revealing the same $1/f^2$ dependence. In future experiments it may be possible to greatly suppress the effect of laser intensity noise by either employing a noise eater to directly reduce it, or measuring it simultaneously with the back-scatter measurements and subtracting it in post-processing.

To determine the frequency noise suppression achieved by the back-scatter technique, we apply a 16 kHz modulation to the laser frequency via it's internal PZT, and characterise the power spectrum at this frequency as a function of laser detuning $\Delta$ as shown in the inset of Fig.~\ref{FreqNoise}. As expected, the observed signal is minimised at zero detuning, with 27~dB of frequency noise suppression evident compared to the far detuned case. With this level of suppression the laser frequency noise, evident as the structural features in the PDH error-signal trace in Fig.~\ref{FreqNoise}, lies more than 10~dB beneath the $1/f^2$ noise on the back-scatter measurement, and is therefore negligible. The suppression achieved here compares favourably to the 20~dB of suppression reported in Ref.~\cite{Lu-2011} using a cooled and stabilised reference interferometer.

\begin{figure}
\begin{center}
\includegraphics[width=0.47\textwidth]{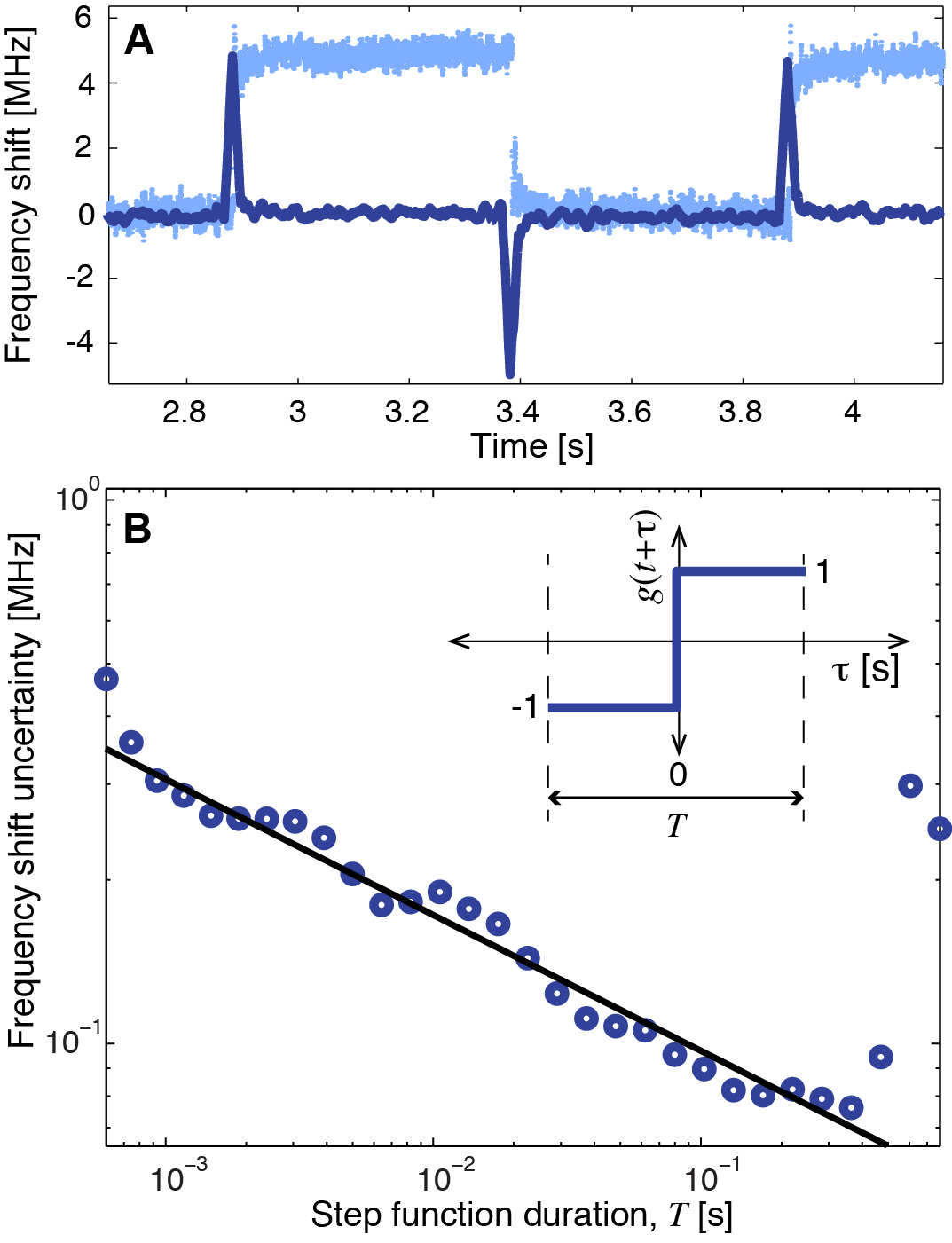}
\caption{{\bf Absolute noise floor of back-scatter measurement.} {\bf A}: Frequency shift  as function of time. Light blue curve: raw frequency shift data $\nu(t)$. Dark blue curve: instantaneous frequency shift $\delta \nu(t)$ obtained using a step function of duration $T = 30$~ms. {\bf B}: Frequency shift uncertainty as a function of step function duration.  Black line: fit to frequency shift uncertainty with function $\Delta \nu = A t^{-1/4}$, $A=54$~kHz~s$^{1/4}$. Inset: step function $g(t+\tau)$ used to perform cross-correlation.}
\label{PwrCurve}
\end{center}
\end{figure}

To demonstrate the absolute noise performance of the back-scatter measurement, a second experiment was conducted, where a square-wave was applied to the AFM position, simulating regular binding events of nanoparticles or biomolecules. As expected, the back-scattered intensity displayed the same square-wave feature (shown in Fig.~\ref{PwrCurve}A). A two step process was performed to calibrate the back-scattered intensity in terms of whispering gallery mode resonance frequency shift. First, the response of the laser frequency to a voltage applied to its internal PZT was characterised. The PDH error signal fed back to the laser then provided the frequency shift of the optical mode as a function of the position of the AFM tip, which could be related back to the observed back-scattered intensity. As can be seen in Fig.~\ref{PwrCurve}A, when present, the AFM caused a 5~MHz frequency shift, with raw frequency noise at a level of 400~kHz. It is important to note that this raw frequency noise depends greatly on the choice of filtering in the data acquisition and should not be treated with real significance.

To determine the frequency noise floor in a meaningful way a cross-correlation was performed between the raw data and a step function. The time-domain filter $\delta \nu (t) = \frac{2}{T}\int_{-T/2}^{T/2} \nu(t) g(t+\tau) d \tau$ was applied to the raw frequency shift data $\nu(t)$ obtained from the back-scatter measurement, where $g(t+\tau)$ is the step-function shown in the inset of Fig.~\ref{PwrCurve}B and $T$ is its duration. As shown in Fig.~\ref{PwrCurve}A (dark blue trace), this converts the frequency shift at time $t$ from the total cumulative frequency shift experienced prior to that time (light blue trace), to the frequency shift $\delta \nu(t)$ due to a binding event at time $t$ given that no other binding events occur within the time window $t \pm T/2$. The standard deviation of $\delta \nu(t)$ over the time window where the AFM is stationary provides the frequency noise floor of the measurement. This noise floor depends on the total duration of the step function, improving as the duration increases and more data is included in the cross-correlation, as shown in Fig.~\ref{PwrCurve}B. A minimum noise floor of 76~kHz is obtained when the duration of the step function is just under 0.5~s, half the period of the square-wave. A dramatic and spurious degradation in the noise floor is evident once the duration exceeds this time and begins to sample the next simulated binding event. Note that in a real scenario involving uncontrolled detection of nanoparticles or biomolecules, the optimal duration of the step function would be dictated by the magnitude of low frequency noise.

In conclusion, we have proposed and demonstrated a new form of self-referenced whispering gallery mode sensing. Unlike other approaches the technique both allows sensing of frequency shifts smaller than the optical linewidth, and does not require any sophisticated frequency stabilisation or characterisation techniques. Rather, it simply requires direct detection of the back-scattered intensity from a conventional whispering gallery mode sensor. We show theoretically, that the quantum shot noise floor of this approach is equal to that achieved from frequency shift measurement of an equivalent single mode optical cavity, and that the thermorefractive noise limit is halved. Furthermore, using an AFM tip to simulate biomolecule binding events, we demonstrate that the technique suppresses laser frequency noise by 27~dB, and out-completes direct frequency shift measurements over the frequency band relevant to biosensing. An absolute sensitivity of 76~kHz is achieved, to our knowledge surpassing all previous whispering gallery mode sensors. The technique is applicable not only to conventional whispering gallery mode sensing configurations, but also to enhanced configurations relying on plasmonic hot-spots or the thermo-optic effect.

\emph{Acknowledgements} This research was funded by the Australian Research Council grant DP0987146. Device fabrication was undertaken within the Queensland Node of the Australian Nanofabrication Facility. 

\clearpage

\begin{center}
{\bf Supplementary information: Back-scatter based whispering gallery mode sensing}
\end{center}

\author{Joachim Knittel} \affiliation{School of Mathematics and Physics, University of Queensland, St Lucia, QLD 4072, Australia} 

\author{Jon D. Swaim} \affiliation{School of Mathematics and Physics, University of Queensland, St Lucia, QLD 4072, Australia} 

\author{David L. McAuslan} \affiliation{Centre for Engineered Quantum Systems, School of Mathematics and Physics, University of Queensland, St Lucia, Brisbane, QLD 4072, Australia}

\author{George A. Brawley} \affiliation{Centre for Engineered Quantum Systems, School of Mathematics and Physics, University of Queensland, St Lucia, Brisbane, QLD 4072, Australia}

\author{Warwick P. Bowen} \affiliation{Centre for Engineered Quantum Systems, School of Mathematics and Physics, University of Queensland, St Lucia, Brisbane, QLD 4072, Australia}

\begin{center}
\quote{ \small
Here we derive the sensitivity of back-scatter based whispering gallery mode sensing, and compare it directly to typical dispersive sensing with an optical cavity. The quantum shot noise limit of the two kinds of sensing is found to be identical. However, in contrast to typical dispersive sensing, back-scatter sensing is found to be, in principle, immune to frequency noise.  Furthermore, the thermorefractive noise is found to be exactly half that of typical sensing. 
}
\end{center}

\section{A brief explanation of the approach}

In what follows we use a standard quantum optics approach to determining the sensitivity limits of WGM sensors (see Ref.~\cite{Gardiner} for an overview of the method). This approach is nice since it naturally includes, in a formal way, the shot noise due to quantisation of the optical field. However, the approach can be understood classically.  Since the entire system is linear, an equivalent result would be obtained by neglecting quantum noise terms (terms like $\delta \tilde a$), thinking of $\alpha_j = \langle a_j \rangle$ as being the appropriately scaled electric field amplitude in mode $j$, and adding  shot noise phenomenologically at the end of the derivation.

In the derivation we make the approximation that the sideband frequency $\omega$, total scattering rate between counter propagating modes $g_0 + g_{\rm sig}$, cavity detuning $\Delta$, and thermorefractive frequency noise are all small compared to the optical linewidth $\gamma$. This approximation is accurate for the vast majority of WGM biosensing experiments in water since the optical linewidth is typically above 10~MHz, and each of the terms listed above are typically less that 1~MHz. The theory could, in principle, quite easily be derived without these approximations. However, the mathematics would become much more complex and the end result less enlightening. 

\section{Deriving the Hamiltonian}

Consider two counter-propagating degenerate WGM resonances, a clock-wise mode (C) and anti-clockwise mode (A), with respective modeshapes $U_C ({\bf r}) = V(r,\theta) e^{i m \phi}$ and $U_A ({\bf r}) = V(r,\theta) e^{-i m \phi}$, where $r$, $\theta$, and $\phi$ are the usual cylindrical co-ordinates, and $m$ is the azimuthal mode number.  The bare Hamiltonian of the system is 
\begin{equation}
H_0 = \hbar \Omega \left (\hat{a}_A^\dagger \hat{a}_A +\hat{a}_C^\dagger \hat{a}_C \right )
\end{equation}
where we have neglected the constant term due to the zero-point energy of the system.

The total Hamiltonian of the system contains both the bare Hamiltonian and terms $\mathcal{V}$ due to the energy of polarizable particles within the electric field $\hat E$ each mode. Defining the modeshapes to be normalised we have $\int |U_A|^2 d{\bf V} = \int |U_C|^2 d{\bf V} = 1$; and since $\int |e^{\pm i m \phi}|^2 d \phi = 2 \pi$, $\int |V(r,\theta)|^2 dr d \theta =1/2 \pi$. Assuming no other resonant modes are nearby, the electric field within the WGM resonator can be approximated by a sum of the electric fields of each of these modes
\begin{eqnarray}
\hat{E} &=& \mathcal{E} e^{i \Omega t} \left ( \hat{a}^\dagger_A U_A^*  ({\bf r}) + \hat{a}^\dagger_C U^*_C  ({\bf r}) \right )+ h.c.\\
&=& \mathcal{E} V^*(r, \theta) e^{i \Omega t} \left ( \hat{a}^\dagger_A e^{i m \phi}   + \hat{a}^\dagger_C e^{- i m \phi} \right ) + h.c.
\end{eqnarray}
where $\hat a_j$ is the annihilation operator of mode $j$ normalised such that $\langle \hat{a}_j^\dagger \hat{a}_j \rangle = \langle \hat {n}_j \rangle$ and $[\hat{a}_j, \hat{a}_j^\dagger ] =1$, and $\mathcal{E}$ is the quantised field strength (i.e. the zero point electric field magnitude in each mode) 
\begin{equation}
\mathcal{E} = \sqrt{\frac{\hbar \Omega}{2 \epsilon_0 V}},
\end{equation}
where, since the two modes are symmetric apart from propagation direction, with degenerate frequency and volume, their quantised field strengths are identical.

We wish to consider microscopic fluctuations in the polarizability of the WGM structure. Therefore, we introduce the polarizability density $\rho_\alpha ({\bf r})$ which describes the capacity of an electric field to induce a change in polarization around the point ${\bf r}$. The polarization density is then just
\begin{eqnarray}
\rho_{p}({\rm r}) &=& \epsilon_0 \rho_\alpha ({\bf r}) \hat{E}({\bf r})\\
&=& \epsilon_0 \mathcal{E} \rho_\alpha ({\bf r}) \left [ V^*(r, \theta) e^{i \Omega t} \left ( \hat{a}^\dagger_A e^{i m \phi}   + \hat{a}^\dagger_C e^{- i m \phi} \right ) +  V(r, \theta) e^{-i \Omega t} \left ( \hat{a}_A e^{-i m \phi}   + \hat{a}_C e^{i m \phi} \right ) \right ].
\end{eqnarray}
Since for a single dipole, the polarization energy is $\mathcal{V} = - {\rm p} \cdot \hat{E} =  - p \hat{E}$, the polarization energy density is
\begin{eqnarray}
\rho_{\mathcal{V}}({\rm r}) &=& - \rho_p({\bf r}) \hat{E}\\
&=& - \epsilon_0 \mathcal{E}^2 \rho_\alpha ({\bf r}) \left [ V^*(r, \theta) e^{i \Omega t} \left ( \hat{a}^\dagger_A e^{i m \phi}   + \hat{a}^\dagger_C e^{- i m \phi} \right ) +  V(r, \theta) e^{-i \Omega t} \left ( \hat{a}_A e^{-i m \phi}   + \hat{a}_C e^{i m \phi} \right ) \right ]^2\\
&=& - \epsilon_0 \mathcal{E}^2 \rho_\alpha ({\bf r}) |V(r, \theta)|^2  \left [ \hat{a}^\dagger_A \hat{a}_A + \hat{a}^\dagger_C \hat{a}_C + \hat{a}^\dagger_A \hat{a}_C e^{2 i m \phi} + \hat{a}_A \hat{a}^\dagger_C e^{-2 i m \phi} \right ].
\end{eqnarray}
Where we've made the usual rotating wave approximation, neglecting fast oscillating terms in time. The polarization energy is then 
\begin{eqnarray}
\mathcal{V} & = & \int \rho_{\mathcal{V}}({\rm r})  d {\bf V}\\
&=& - \epsilon_0 \mathcal{E}^2 \left [ \int \rho_\alpha ({\bf r}) |V(r, \theta)|^2 d{\bf V} \left (\hat{a}^\dagger_A \hat{a}_A + \hat{a}^\dagger_C \hat{a}_C \right) + \int \rho_\alpha ({\bf r}) |V(r, \theta)|^2 e^{2 i m \phi} d{\bf V} \hat{a}^\dagger_A \hat{a}_C  + \int \rho_\alpha ({\bf r}) |V(r, \theta)|^2 e^{-2 i m \phi} d{\bf V} \hat{a}_A \hat{a}^\dagger_C  \right ]\\
&=&-  \hbar g_{\rm self} \left (\hat{a}^\dagger_A \hat{a}_A + \hat{a}^\dagger_C \hat{a}_C \right)  - \hbar g_{\rm cross} \hat{a}^\dagger_A \hat{a}_C - \hbar g_{\rm cross}^* \hat{a}_A \hat{a}^\dagger_C
\end{eqnarray}
where 
\begin{eqnarray}
\hbar g_{\rm self}&=&\epsilon_0 \mathcal{E}^2 \int \rho_\alpha ({\bf r}) |V(r, \theta)|^2 d{\bf V} \label{gself}\\
\hbar g_{\rm cross} &=& \epsilon_0 \mathcal{E}^2  \int \rho_\alpha ({\bf r}) |V(r, \theta)|^2 e^{2 i m \phi} d{\bf V}, \label{gcross}
\end{eqnarray}
respectively represent the polarization induced self-energy of modes $A$ and $C$, the interaction energy between them.

We consider three different contributions to the polarizability density: the nanoparticle that we seek to detect with point-like polarizability $\alpha_{\rm sig}$ located at ${\bf r}_{\rm sig}$, a static defect scattering center on the WGM with point-like polarizability $\alpha_{0}$ located at ${\bf r}_0$, and a spatially and temporally fluctuating polarizability density $\rho_{\alpha, \rm therm} ({\rm r}, t)$. The  total polarizability density is then
\begin{equation}
\rho_\alpha ({\bf r}) = \alpha_{\rm sig} \delta({\bf r}_{\rm sig}) +  \alpha_{\rm 0} \delta({\bf r}_{\rm 0}) + \rho_{\alpha, \rm therm} ({\bf r}, t).
\end{equation}
Substituting this into Eq.~(\ref{gself}) above we find
\begin{eqnarray}
\hbar g_{\rm self}&=&  \epsilon_0 \mathcal{E}^2 \int \left ( \alpha_{\rm sig} \delta({\bf r}_{\rm sig}) +  \alpha_{\rm 0} \delta({\bf r}_{\rm 0}) + \rho_{\alpha, \rm therm} ({\bf r}, t) \right )|V(r, \theta)|^2 d{\bf V}\\
&=& \epsilon_0 \mathcal{E}^2  \left (\alpha_{\rm sig} |V(r_{\rm sig}, \theta_{\rm sig})|^2 + \alpha_{0} |V(r_0, \theta_0)|^2 + \int \rho_{\alpha, \rm therm} ({\bf r}, t) |V(r, \theta)|^2 d{\bf V} \right )\\
&=& \hbar \left ( g_{\rm sig} + g_0 + g_{\rm therm}^{\rm self} \right )
\end{eqnarray}
Similarly, substituting into Eq.~(\ref{gcross}) we find 
\begin{eqnarray}
\hbar g_{\rm cross}&=& \epsilon_0 \mathcal{E}^2 \int \left ( \alpha_{\rm sig} \delta({\bf r}_{\rm sig}) +  \alpha_{\rm 0} \delta({\bf r}_{\rm 0}) + \rho_{\alpha, \rm therm} ({\bf r}, t) \right )|V(r, \theta)|^2 e^{2 i m\phi} d{\bf V}\\
&=& \epsilon_0 \mathcal{E}^2  \left (\alpha_{\rm sig} |V(r_{\rm sig}, \theta_{\rm sig})|^2 e^{2 i m\phi_{\rm sig}}  + \alpha_{0} |V(r_0, \theta_0)|^2 e^{2 i m\phi_0}  + \int \rho_{\alpha, \rm therm} ({\bf r}, t) |V(r, \theta)|^2 e^{2 i m\phi} d{\bf V} \right )\\
&=& \hbar \left (g_{\rm sig}  e^{2 i m\phi_{\rm sig}}  + g_0 e^{2 i m\phi_0} + g_{\rm therm}^{\rm cross} \right )
\end{eqnarray}

The modification to the Hamiltonian $\mathcal V$ due to polarizable particles is therefore
\begin{equation}
\mathcal{V} = - \hbar \left ( g_{\rm sig} + g_0 + g_{\rm therm}^{\rm self} \right ) \left (\hat{a}^\dagger_A \hat{a}_A + \hat{a}^\dagger_C \hat{a}_C \right) - \hbar \left ( g_{\rm sig}  e^{2 i m\phi_{\rm sig}}  + g_0 e^{2 i m\phi_0} + g_{\rm therm}^{\rm cross}  \right ) \hat{a}^\dagger_A \hat{a}_C  - \hbar \left ( g_{\rm sig}  e^{-2 i m\phi_{\rm sig}}  + g_0 e^{-2 i m\phi_0} + g_{\rm therm}^{{\rm cross}~*}  \right ) \hat{a}_A \hat{a}_C^\dagger 
\end{equation}
with the total Hamiltonian being 
\begin{equation}
H=H_0+\mathcal{V} 
\end{equation}
We see that each of the nanoparticle, defect centre, and thermorefractive noise introduce both a shift in the WGM resonance frequencies, and a coupling rate between them.

We can simplify the problem, by moving into a rotating frame at  frequency $\Omega - \Delta_0$, where $\Delta_0$ is the detuning of the laser field from the bare optical resonance. To do this, we apply the unitary $\hat U = e^{-i(\Omega + \Delta_0)t}$ such that $\tilde{H} = \hat U^{\dagger} \hat H \hat U - \hbar (\Omega- \Delta_0)$; where the $\tilde{}$ denotes throughout a Hamiltonian or operator in the the rotating frame. We then arrive at
\begin{equation}
\tilde{H} = - \hbar \left ( g_{\rm sig} + g_{\rm therm}^{\rm self} + \Delta \right ) \left (\tilde{a}^\dagger_A \tilde{a}_A + \tilde{a}^\dagger_C \tilde{a}_C \right) - \hbar \left ( g_{\rm sig}  e^{2 i m\phi_{\rm sig}}  + g_0 e^{2 i m\phi_0} + g_{\rm therm}^{\rm cross}  \right ) \tilde{a}^\dagger_A \tilde{a}_C  - \hbar \left ( g_{\rm sig}  e^{-2 i m\phi_{\rm sig}}  + g_0 e^{-2 i m\phi_0} + g_{\rm therm}^{{\rm cross}~*}  \right ) \tilde{a}_A \tilde{a}_C^\dagger,  \label{H}
\end{equation}
with $\tilde{a}=\hat{a}e^{-i(\Omega- \Delta)t}$, and $\Delta = \Delta_0 + g_0$ being the modified cavity defining due to the scattering centre $g_0$.

\subsection{Variances of fluctuating energy terms}

The energy terms due to thermorefractive noise $g_{\rm therm}^{\rm self}$ and $g_{\rm therm}^{\rm cross}$ are both zero-mean noise terms. Later on in this document it will be useful to know the relationship between their variances. 

the variance of $g_{\rm therm}^{\rm self}$  is
\begin{eqnarray}
\hbar^2 \left \langle {g_{\rm therm}^{\rm self}}^2\right \rangle &=& \epsilon_0^2 \mathcal{E}^4 \left \langle \int \rho_{\alpha, \rm therm} ({\bf r}, t) |V(r, \theta)|^2 d{\bf V}  \int \rho_{\alpha, \rm therm} ({\bf r}', t) |V(r', \theta')|^2  d{\bf V}'  \right \rangle \\
&=& \epsilon_0^2 \mathcal{E}^4   \iint \left  \langle \rho_{\alpha, \rm therm} ({\bf r}, t)  \rho_{\alpha, \rm therm} ({\bf r}', t)  \right \rangle |V(r, \theta)|^2  |V(r', \theta')|^2  d{\bf V}  d{\bf V}' \\
&=& \epsilon_0^2 \mathcal{E}^4  \int  \left \langle \rho_{\alpha, \rm therm} ({\bf r}, t)^2 \right \rangle  |V(r, \theta)|^4 d{\bf V}\\
&=& \frac{\epsilon_0^2 \mathcal{E}^4}{2 \pi}  \iint  \left \langle \rho_{\alpha, \rm therm} (r,\theta, t)^2 \right \rangle  |V(r, \theta)|^4 dr d\theta
\end{eqnarray}
where we've used the usual approximation that thermorefractive noise is delta-correlated, $\langle \rho_{\alpha, \rm therm} ({\bf r}, t)  \rho_{\alpha, \rm therm} ({\bf r}', t)  \rangle = \langle \rho_{\alpha, \rm therm} ({\bf r}, t)^2 \rangle \delta_{{\rm r} = {\rm r}'}$, and, assuming a homogeneous material have taken the average thermorefractive noise to have no $\phi$ dependence.

Unlike $g_{\rm therm}^{\rm self}$, $g_{\rm therm}^{\rm cross}$ can in general be complex.  We therefore break it into real and imaginary parts $g_{\rm therm}^{\rm cross} = R_{\rm therm}^{\rm cross} + i  I_{\rm therm}^{\rm cross}$  and calculate the variance of each independently.
\begin{eqnarray}
\hbar R_{\rm therm}^{\rm cross} &=& \int \rho_{\alpha, \rm therm} ({\bf r}, t) |V(r, \theta)|^2 \cos{2 m\phi} d{\bf V}\\
\hbar I_{\rm therm}^{\rm cross} &=& \int \rho_{\alpha, \rm therm} ({\bf r}, t) |V(r, \theta)|^2 \sin{2 m\phi} d{\bf V}
\end{eqnarray}

The real variance is then
\begin{eqnarray}
\hbar^2 \left \langle {R_{\rm therm}^{\rm cross}}^2\right \rangle &=&  \epsilon_0^2 \mathcal{E}^4 \left \langle \int \rho_{\alpha, \rm therm} ({\bf r}, t) |V(r, \theta)|^2 \cos{2 m\phi} d{\bf V} \int \rho_{\alpha, \rm therm} ({\bf r}', t) |V(r', \theta')|^2 \cos{2 m\phi'} d{\bf V}'  \right \rangle\\
&=& \epsilon_0^2 \mathcal{E}^4 \iint \left \langle  \rho_{\alpha, \rm therm} ({\bf r}, t)  \rho_{\alpha, \rm therm} ({\bf r}', t)  \right \rangle |V(r, \theta)|^2 |V(r', \theta')|^2 \cos{2 m\phi}  \cos{2 m\phi'} d{\bf V} d{\bf V}'  \\
&=& \epsilon_0^2 \mathcal{E}^4 \int \left \langle  \rho_{\alpha, \rm therm} ({\bf r}, t)^2 \right \rangle |V(r, \theta)|^4 \cos^2{2 m\phi}   d{\bf V} \\
&=& \frac{\epsilon_0^2 \mathcal{E}^4}{2} \iint \left \langle  \rho_{\alpha, \rm therm} ({\bf r}, t)^2 \right \rangle |V(r, \theta)|^4 dr d \theta \int 1+ \cos{4 m\phi}  d \phi\\
&=& \frac{\epsilon_0^2 \mathcal{E}^4}{4 \pi} \iint \left \langle  \rho_{\alpha, \rm therm} (r, \theta, t)^2 \right \rangle |V(r, \theta)|^4 dr d \theta\\
&=& \frac{1}{2} \hbar^2 \left \langle {g_{\rm therm}^{\rm self}}^2\right \rangle
\end{eqnarray}
Similarly for the imaginary term
\begin{eqnarray}
\hbar^2 \left \langle {I_{\rm therm}^{\rm cross}}^2\right \rangle &=&  \epsilon_0^2 \mathcal{E}^4 \left \langle \int \rho_{\alpha, \rm therm} ({\bf r}, t) |V(r, \theta)|^2 \sin{2 m\phi} d{\bf V} \int \rho_{\alpha, \rm therm} ({\bf r}', t) |V(r', \theta')|^2 \cos{2 m\phi'} d{\bf V}'  \right \rangle\\
&=& \epsilon_0^2 \mathcal{E}^4 \iint \left \langle  \rho_{\alpha, \rm therm} ({\bf r}, t)  \rho_{\alpha, \rm therm} ({\bf r}', t)  \right \rangle |V(r, \theta)|^2 |V(r', \theta')|^2 \sin{2 m\phi}  \sin{2 m\phi'} d{\bf V} d{\bf V}'  \\
&=& \epsilon_0^2 \mathcal{E}^4 \int \left \langle  \rho_{\alpha, \rm therm} ({\bf r}, t)^2 \right \rangle |V(r, \theta)|^4 \sin^2{2 m\phi}   d{\bf V} \\
&=& \frac{\epsilon_0^2 \mathcal{E}^4}{2} \iint \left \langle  \rho_{\alpha, \rm therm} ({\bf r}, t)^2 \right \rangle |V(r, \theta)|^4 dr d \theta \int 1- \cos{4 m\phi}  d \phi\\
&=& \frac{\epsilon_0^2 \mathcal{E}^4}{4 \pi} \iint \left \langle  \rho_{\alpha, \rm therm} (r, \theta, t)^2 \right \rangle |V(r, \theta)|^4 dr d \theta\\
&=& \frac{1}{2} \hbar^2 \left \langle {g_{\rm therm}^{\rm self}}^2\right \rangle
\end{eqnarray}

Hence we see that
\begin{equation}
\left \langle {R_{\rm therm}^{\rm cross}}^2\right \rangle =  \left \langle {I_{\rm therm}^{\rm cross}}^2\right \rangle = \frac{1}{2} \left \langle {g_{\rm therm}^{\rm self}}^2\right \rangle \label{therm_relation}
\end{equation}

\section{Direct measurement of the phase shift on a single mode as a benchmark}

The usual approach to dispersive measurement in an optical resonator is to excite the resonator on resonance, and directly measure the phase shift introduced on the field exiting the resonator. This approach has been shown, in principle, to be optimal, reaching the quantum limit due to shot noise. Here, prior to considering the full measurement protocol, we consider this simplified case to derive the quantum limit on measurements of the signal scattering rate $g_{\rm sig}$ (or equivalently, the polarizability of the nanoparticle). The Hamiltonian is then found by dropping all of the coupling terms from Eq.~(\ref{H}), and considering only one of the two counter-propagating modes, dropping the redundant subscript.
\begin{eqnarray}
\tilde{H}_{\rm QNL} &=& - \hbar \left ( g_{\rm sig} +  g_{\rm therm}^{\rm self} + \Delta \right ) \tilde{a}^\dagger \tilde{a}\\
&=& - \hbar \left ( g_{\rm sig} + g_{\rm therm}^{\rm self} + \Delta \right ) \tilde{a}^\dagger \tilde{a}
\end{eqnarray}

The equation of motion for the operator $\tilde a$ may then be determined from the quantum Langevin equation
\begin{eqnarray}
\dot{\tilde a}(t) &=& \frac{1}{i \hbar} \left [\tilde a(t), \tilde H_{\rm QNL} \right ] - \gamma(t) \tilde a(t) + \sqrt{2 \gamma_{\rm in}(t)} \tilde a_{\rm in}(t) + \sqrt{2 \gamma_{l}} \delta \tilde a_{l}(t) \\
&=&  i \left ( g_{\rm sig} + g_{\rm therm}^{\rm self} + \Delta \right ) \tilde{a}(t) - \gamma(t) \tilde a(t) + \sqrt{2 \gamma_{\rm in}(t)} \tilde a_{\rm in}(t) + \sqrt{2 \gamma_{l}} \delta\tilde a_{l}(t) \\
&=&\left [- \gamma(t)  + i \left ( g_{\rm sig} + g_{\rm therm}^{\rm self} + \Delta \right ) \right ] \tilde{a}(t) + \sqrt{2 \gamma_{\rm in}(t)} \tilde a_{\rm in}(t) + \sqrt{2 \gamma_{l}} \delta\tilde a_{l}(t),\label{eqmoQNL}
\end{eqnarray}
where we have used the Boson commutation relation $[\tilde a,\tilde a^\dagger]=1$, $\gamma(t) = \gamma_{\rm in}(t) + \gamma_l (t)$ is the total decay rate of the optical resonator, $\gamma_{\rm in}(t)$ is the input coupling rate with explicit time dependence included to allow noise in this parameter to be included in the model,  $\gamma_{\rm in}(t)$ is the loss rate of the resonator, and $\tilde a_{\rm in}$ and  $\delta \tilde a_{l}$ are respectively the bright field entering the resonator through in the input coupler and vacuum noise entering through loss, with $\tilde n = \tilde a_{\rm in}^\dagger \tilde a_{\rm in}$ being the incident optical intensity in units of photons per second.

\subsubsection{Input noise terms}

Each of the annihilation operators representing  input fields may be broken down into amplitude $\tilde X$ and phase $\tilde Y$ quadratures as
\begin{equation}
\tilde a = \frac{\tilde X + i \tilde Y}{2}.
\end{equation}
The Boson commutation relation, results in a commutation relation between $\tilde X (\omega)$ and $\tilde Y (\omega)$, $[\tilde X (\omega), \tilde Y (\omega)] = 2 i$, such that their is an uncertainty principle relating the two quadratures $\Delta \tilde X \Delta \tilde Y \ge 1$. It is this uncertainty principle which, from the perspective of quantum mechanics, results in the shot noise limit to sensing. If the input laser is coherent at frequency $\omega$, $\langle \delta \tilde X_{\rm in}(\omega)^2 \rangle = \langle \delta \tilde Y_{\rm in}(\omega)^2 \rangle = 1$. However, it is always the case at low frequencies of interest to biosensing experiments, that classical noise sources enter, both on the amplitude of the light and on the phase.  The noise on these input fields can then be broken into a classical component and a quantum component with unity variance, eg. $\delta \tilde X_{\rm in} (\omega) = \delta \tilde X_q (\omega) + \delta \tilde X_c (\omega)$, where $\langle \delta X_q^2\rangle =1$. Taking the variance of each noise term, and using the fact that the quantum and classical noise is uncorrelated, we find
\begin{eqnarray}
\left \langle \delta \tilde X_{\rm in} (\omega)^2 \right \rangle &=& \left \langle \delta \tilde X_q (\omega)^2 \right \rangle +\left \langle \delta \tilde X_c (\omega)^2 \right \rangle  = 1 + \left \langle \delta \tilde X_c (\omega)^2 \right \rangle \\
\left \langle \delta \tilde Y_{\rm in} (\omega)^2 \right \rangle &=& \left \langle \delta \tilde Y_q (\omega)^2 \right \rangle +\left \langle \delta \tilde Y_c (\omega)^2 \right \rangle  = 1 + \left \langle \delta \tilde Y_c (\omega)^2 \right \rangle
\end{eqnarray}
The classical phase noise can be conveniently reexpressed in terms of absolute phase noise, i.e. $\langle \delta \tilde Y_c (\omega)^2  \rangle = V_\zeta (\omega) \alpha_{\rm in}^2$, where $V_\zeta (\omega)$ is the variance of classical phase noise, scaled by the incident optical intensity $\alpha_{\rm in}^2$, since the absolute displacement in optical phase space is amplified by the coherent amplitude of the field. Similarly, the classical amplitude noise may be reexpressed as a relative amplitude noise  $\langle \delta \tilde X_c (\omega)^2  \rangle = V_{\rm RIN} (\omega) \alpha_{\rm in}^2$. It is natural to do this because the variance of classical noise in both amplitude and phase scales linearly with intensity, whilst, due to the injection of vacuum noise when the laser is attenuated for any reason, the quantum noise variance remains constant. $V_\zeta (\omega)$ and $V_{\rm RIN} (\omega)$ are therefore intensity independent noise parameters. We then have
\begin{eqnarray}
\left \langle \delta \tilde X_{\rm in} (\omega)^2 \right \rangle &=& 1 + V_{\rm RIN} (\omega) \alpha_{\rm in}^2 \label{VXin} \\ 
\left \langle \delta \tilde Y_{\rm in} (\omega)^2 \right \rangle &=& 1 + V_{\zeta} (\omega) \alpha_{\rm in}^2 \label{VYin}.
\end{eqnarray}

Any fluctuation in optical path length, such as these due to thermal fluctuations and mechanical vibrations, will introduce phase (or equivalently frequency) noise into the sensing apparatus. By contrast, the introduction of amplitude noise requires fluctuations in optical attenuation which are much less common. Furthermore, optical intensity fluctuations are directly measured with a photodiode, and can therefore by compensated for in measured data, or suppressed through fed back/forward as performed in common "noise eater" arrangements. Consequently, phase and frequency noise tend to be both larger in magnitude and more difficult to suppress in optical sensors \footnote{A noise eater will usually work by tapping some of the laser light out of the system directly measuring it to obtain the fluctuating intensity noise, and using an amplitude modulator either in feedforward or feedback configuration to subtract this noise from the laser. In principle, with such a system, it is possible to reduce the classical noise variance such that it equals the quantum noise variance, ie $V_{\rm RIN} (\omega) \alpha_{\rm in}^2=1$. Commercial devices  operate close to this limit in the frequency band of interest to biosensing ($<$~1 MHz).}

\subsection{Solving the mean field}

The first step to solving the problem is to calculate the mean fields in the resonator and leaving it back through the input coupler. We expand each of the time varying terms in Eq.~(\ref{eqmoQNL}) into their mean values and noise fluctuations ($\tilde a = \langle \tilde a \rangle + \delta \tilde a = \alpha + \delta \tilde a$, $\gamma = \langle \gamma \rangle + \delta \gamma = \bar \gamma + \delta \gamma$), take the expectation value, and linearize by neglecting noise product terms to get
\begin{equation}
\dot{\alpha}(t) = \left (- \bar \gamma  + i  (g_{\rm sig}+ \Delta)  \right ) \alpha(t) + \sqrt{2 \bar \gamma_{\rm in}} \alpha_{\rm in}(t) ,
\end{equation}
where, since $g_{\rm therm}^{\rm self}$ is zero mean thermorefractive noise $\langle g_{\rm therm}^{\rm self} \rangle=0$. 
Choosing the input field mean amplitude to be stationary and taking the steady state ($ \alpha_{\rm in}(t)= \alpha_{\rm in}$, $\dot{\alpha}(t)=0$), we get
\begin{equation}
0 = \left (- \bar \gamma  + i  ( g_{\rm sig} + \Delta)  \right ) \alpha + \sqrt{2 \bar \gamma_{\rm in}} \alpha_{\rm in}
\end{equation}
where henceforth $\alpha$ is implicitly taken to be the steady state intracavity intensity. So that
\begin{equation}
\alpha  = \frac{ \sqrt{2 \bar \gamma_{\rm in}} \alpha_{\rm in}}{ \bar \gamma  - i (g_{\rm sig}+\Delta) }  \label{alphanoapproxQNL}
\end{equation}
The limit relevant to high precision sensing experiments is the limit where both the frequency shift due to the nanoparticle $g_{\rm sig}$ and the cavity detuning are small compared to the cavity linewidth $\bar \gamma$. Therefore, it is reasonable to take the limit $\{ g_{\rm sig}, \Delta \} \ll \bar \gamma$. In this case, it is straightforward to show that Eq.~(\ref{alphanoapproxQNL}) may be well approximated as 
\begin{equation}
\alpha  =  \frac{ \sqrt{2 \bar \gamma_{\rm in}}}{\bar \gamma^2} 
\left ( \bar \gamma  + i ( g_{\rm sig}+ \Delta) \right )  \alpha_{\rm in}  \label{alphaQNL}
\end{equation}

The mean amplitude of the output field may be calculated using the input-output formalism with
\begin{equation}
\tilde a_{\rm out} = \tilde a_{\rm in} - \sqrt{2 \gamma_{\rm in}(t)} \tilde a.
\end{equation}
Since $\gamma_{\rm in}$ varies with time but the amplitude of variation is small compared with the coupling rate itself, the square-root can be expanded as $\sqrt{2 \gamma_{\rm in}(t)} \approx \sqrt{2 \bar \gamma_{\rm in}} + \delta \gamma_{\rm in}(t)/\sqrt{2 \bar \gamma_{\rm in}}$, so that
\begin{equation}
\tilde a_{\rm out} = \tilde a_{\rm in} - \left [\sqrt{2 \bar \gamma_{\rm in}} + \frac{\delta \gamma_{\rm in}}{\sqrt{2 \bar \gamma_{\rm in}}} \right ] \tilde a.
\end{equation}
This results in mean and fluctuation input output relations
\begin{eqnarray}
\alpha_{\rm out} &=& \alpha_{\rm in} - \sqrt{2 \bar \gamma_{\rm in}} \alpha \label{inoutalphaQNL} \\
\delta \tilde a_{\rm out} &=& \delta \tilde a_{\rm in} - \sqrt{2 \bar \gamma_{\rm in}} \delta \tilde a - \frac{\alpha}{\sqrt{2 \bar \gamma_{\rm in}} }\delta \gamma_{\rm in}  \label{inoutdeltaQNL}  .
\end{eqnarray}
 
Using Eqs.~(\ref{alphaQNL})~and~(\ref{inoutalphaQNL}) we then find
\begin{eqnarray}
\alpha_{\rm out} &=& \alpha_{\rm in} -  \frac{ 2 \bar \gamma_{\rm in}}{\bar \gamma^2} 
\left ( \bar \gamma  + i  ( g_{\rm sig}+ \Delta)  \right )  \alpha_{\rm in}  \\
&=& \left [1  - 2 \eta \left (1
+  i  ( g_{\rm sig}'  + \Delta' ) \right ) \right ] \alpha_{\rm in} \label{alphaoutQNL}
\end{eqnarray}
where it will be convenient throughout to normalize rates in terms of the mean cavity decay rate $\bar \gamma$. In these cases, the normalization will generally be denoted with a ${}'$. Here, for example $g_{\rm sig}' = g_{\rm sig}/\bar \gamma$ and ${g_{\rm therm}^{\rm self }}' = g_{\rm therm}^{\rm self }/\bar \gamma$. $\eta = \bar \gamma_{\rm in}/\bar \gamma$ is the escape efficiency of the cavity, i.e. the probability that a photon in the cavity will leave through the input/output coupler. Since we wish to measure the magnitude of $g_{\rm sig}$ and $g_{\rm sig} \ll \bar \gamma$, it is clear from Eq.~(\ref{alphaoutQNL}) that we wish to measure a phase rotation on the output field of the resonator ($g_{\rm sig}$ displaces the amplitude of the field a small amount in a direction in phase space orthogonal to the mean amplitude).

\subsection{Solving for the fluctuating noise}

To solve for the steady-state fluctuations on both intracavity and output fields we make the substitutions $\tilde a = \alpha + \delta \tilde a$ for each annihilation operator in Eq.~\ref{eqmoQNL}, substitute $\gamma(t) = \bar \gamma + \delta \gamma (t)$, and substitute $\sqrt{2 \gamma_{\rm in} (t)} = \sqrt{2 \bar \gamma_{\rm in}} +  \delta \gamma_{\rm in} (t)/\sqrt{2 \bar \gamma_{\rm in}}$. Using Eq.~(\ref{alphaQNL}) the mean terms cancel, neglecting second order fluctuating terms we find
\begin{equation}
\delta \dot{\tilde a}(t) = i ( g_{\rm sig}+ \Delta) \delta \tilde a (t) -\bar \gamma \delta \tilde a (t) - \alpha \delta \gamma_{\rm in}(t) + i \alpha g_{\rm therm}^{\rm self}(t) + \frac{\alpha_{\rm in}}{\sqrt{2 \bar \gamma_{\rm in}}} \delta \gamma_{\rm in}(t) + \sqrt{2 \bar \gamma_{\rm in}} \delta \tilde a_{\rm in}(t) + \sqrt{2 \gamma_l} \delta \tilde a_l (t) 
\end{equation}
This can easily be solved in the frequency domain by taking the Fourier transform, which yields
\begin{eqnarray}
\delta \tilde a (\omega) &=& \left [ \frac{1}{\bar \gamma + i (\omega - g_{\rm sig}- \Delta)} \right ] \left \{ \frac{ \alpha_{\rm in}}{\sqrt{2 \bar \gamma_{\rm in}}} \left [ 1 - \frac{2 \bar \gamma_{\rm in}}{\bar \gamma - i ( g_{\rm sig}+ \Delta) }   \right ] \delta \gamma_{\rm in} (\omega) + i \alpha g_{\rm therm}^{\rm self} (\omega) + \sqrt{2 \bar \gamma_{\rm in}} \delta \tilde a_{\rm in}(\omega) + \sqrt{2 \gamma_l} \delta \tilde a_l (\omega) \right \}\\
&=& \left [ \frac{\bar \gamma - i (\omega - g_{\rm sig}- \Delta)}{\bar \gamma^2} \right ] \left \{ \frac{ \alpha_{\rm in}}{\sqrt{2 \bar \gamma_{\rm in}}} \left [ 1 - \frac{2 \bar \gamma_{\rm in}}{\bar \gamma - i (g_{\rm sig}+ \Delta)}   \right ] \delta \gamma_{\rm in} + i    \alpha_{\rm in}  \frac{ \sqrt{2 \bar \gamma_{\rm in}}}{\bar \gamma} 
 g_{\rm therm}^{\rm self} + \sqrt{2 \bar \gamma_{\rm in}} \delta \tilde a_{\rm in} + \sqrt{2 \gamma_l} \delta \tilde a_l \right \}\\
&=& \left [ \frac{\bar \gamma - i ( \omega-  g_{\rm sig}- \Delta)}{\bar \gamma^2} \right ] \left \{ \frac{ \alpha_{\rm in}}{\sqrt{2 \bar \gamma_{\rm in}}} \left [ 1 - \frac{2 \bar \gamma_{\rm in} (\bar \gamma + i (g_{\rm sig}+ \Delta)) }{\bar \gamma^2}   \right ] \delta \gamma_{\rm in}  + i    \alpha_{\rm in}  \frac{ \sqrt{2 \bar \gamma_{\rm in}}}{\bar \gamma}  g_{\rm therm}^{\rm self} + \sqrt{2 \bar \gamma_{\rm in}} \delta \tilde a_{\rm in}+ \sqrt{2 \gamma_l} \delta \tilde a_l \right \}\\
&=&  \left [ \frac{1 - i (\omega' - g_{\rm sig}' - \Delta')}{\bar \gamma} \right ] \left \{ \frac{ \alpha_{\rm in}}{\sqrt{2 \bar \gamma_{\rm in}}} \left [ 1 - 2 \eta  -2 i  \eta (g_{\rm sig}' + \Delta')  \right ] \delta \gamma_{\rm in} + i    \alpha_{\rm in}  \sqrt{2 \bar \gamma_{\rm in}} {g_{\rm therm}^{\rm self}}'  + \sqrt{2 \bar \gamma_{\rm in}} \delta \tilde a_{\rm in}+ \sqrt{2 \gamma_l} \delta \tilde a_l  \right \},
\end{eqnarray}
where we have substituted in for $\alpha$ from Eq.~(\ref{alphaQNL}), neglected the term involving the product $( g_{\rm sig}+ \Delta)  g_{\rm therm}^{\rm self}$ since this is small compared with $\bar \gamma^2$, and taken $\omega \ll \bar \gamma$. For compactness we've omitted the explicit frequency dependence $\omega$ of the fluctuations here and henceforth.

The output field fluctuations can then be found using Eq.~(\ref{inoutdeltaQNL})
\begin{eqnarray}
\delta \tilde a_{\rm out} (\omega) &=& \delta \tilde a_{\rm in} -  \left [ \frac{1 - i (\omega' - g_{\rm sig}' - \Delta')}{\bar \gamma} \right ] \left \{ \alpha_{\rm in} \left [ 1 - 2 \eta  -2 i  \eta (g_{\rm sig}' + \Delta')  \right ] \delta \gamma_{\rm in} + 2 i  \bar \gamma_{\rm in}  \alpha_{\rm in}  {g_{\rm therm}^{\rm self}}'  + 2 \bar \gamma_{\rm in} \delta \tilde a_{\rm in}+ 2 \sqrt{ \gamma_{\rm in} \gamma_l } \delta \tilde a_l  \right \} \\
&& - \frac{\alpha}{\sqrt{2 \bar \gamma_{\rm in}} }\delta \gamma_{\rm in} \\
&=& \delta \tilde a_{\rm in} -  \left [ \frac{1 - i (\omega' - g_{\rm sig}' - \Delta')}{\bar \gamma} \right ] \left \{ \alpha_{\rm in} \left [ 1 - 2 \eta  -2 i  \eta (g_{\rm sig}' + \Delta')  \right ] \delta \gamma_{\rm in} + 2 i  \bar \gamma_{\rm in}  \alpha_{\rm in}  {g_{\rm therm}^{\rm self}}'  + 2 \bar \gamma_{\rm in} \delta \tilde a_{\rm in}+ 2 \sqrt{ \gamma_{\rm in} \gamma_l } \delta \tilde a_l  \right \} \\
&&- \frac{1}{\bar \gamma}  \left (1  + i ( g_{\rm sig}'+ \Delta') \right )  \alpha_{\rm in} \delta \gamma_{\rm in} \\
&=&   \delta \tilde a_{\rm in} -  \left (1 - i (\omega' - g_{\rm sig}' - \Delta') \right ) \left \{ \alpha_{\rm in} \left [ 1 - 2 \eta  -2 i  \eta (g_{\rm sig}' + \Delta')  \right ] \delta \gamma_{\rm in}' + 2 i  \eta  \alpha_{\rm in}  {g_{\rm therm}^{\rm self}}'  + 2 \eta \delta \tilde a_{\rm in}+ 2\sqrt{\eta (1-\eta)} \delta \tilde a_l  \right \} \\
&& -  \left (1  + i ( g_{\rm sig}'+ \Delta') \right )  \alpha_{\rm in} \delta \gamma_{\rm in}' \\
&=& \left [ 1 - 2 \eta \left (1 - i  (\omega' - g_{\rm sig}' - \Delta')  \right )  \right ] \delta \tilde a_{\rm in} 
-  2\sqrt{\eta (1-\eta)} \left (1 - i (\omega' - g_{\rm sig}' - \Delta') \right ) \delta a_l \\
&& - \alpha_{\rm in} \left [1  + i ( g_{\rm sig}'+ \Delta')  + \left (1 - i (\omega' - g_{\rm sig}' - \Delta') \right )  \left ( 1 - 2 \eta  -2 i  \eta (g_{\rm sig}' + \Delta')  \right ) \right ]  \delta \gamma_{\rm in}' \\
&& - 2 i  \eta  \alpha_{\rm in}   \left (1 - i (\omega' - g_{\rm sig}' - \Delta') \right ) {g_{\rm therm}^{\rm self}}' \label{daoutQNL}
\end{eqnarray}

The output field annihilation operator is given by $\tilde a_{\rm out} = \alpha_{\rm out} + \delta \tilde a_{\rm out}$. As can be seen from Eq.~(\ref{alphaoutQNL}), the signal $g'_{\rm sig}$ is contained entirely within the imaginary part of $\alpha_{\rm out}$. It can be shown from Eq.~(\ref{daoutQNL}) that the noise variance is independent of the measured phase. Consequently, the optimal approach to extracting a signal from $\tilde a_{\rm out}$ is to measure the phase quadrature $\tilde Y_{\rm out}(\omega) = i \left [ \tilde a_{\rm out}^\dagger (-\omega) - \tilde a_{\rm out} (\omega) \right ]$. From Eq.~(\ref{alphaoutQNL}) we have
\begin{equation}
\left \langle \tilde Y_{\rm out} \right \rangle =  - 4 \eta \alpha_{\rm in} \left ( g'_{\rm sig} + \Delta' \right )
\end{equation}
Thus, an estimate of $g'_{\rm sig}$ may be obtained from a measurement of $\tilde Y_{\rm out}$ as
\begin{equation}
{g_{\rm sig}^{\rm est}}' = - \frac{\tilde Y_{\rm out}}{4 \eta \alpha_{\rm in}} - \Delta' = - \frac{ \left \langle \tilde Y_{\rm out} \right \rangle}{4 \eta \alpha_{\rm in}} - \Delta' -\frac{\delta \tilde Y_{\rm out}}{4 \eta \alpha_{\rm in}}
\end{equation}
which in the limit that no noise was present in the measurement, would exactly retrieve $g'_{\rm sig}$. It can be seen from this equation that the variance, or uncertainty, of the estimate is given by
\begin{equation}
\left \langle \left | {\delta g_{\rm sig}^{\rm est}}' \right |^2 \right \rangle_{\rm QNL}= \left \langle \left | {g_{\rm sig}^{\rm est}}' \right |^2 \right \rangle - \left | \left \langle {g_{\rm sig}^{\rm est}}'  \right \rangle \right | ^2 = \frac{\left \langle \left | \delta \tilde Y_{\rm out} \right | ^2 \right \rangle }{16 \eta^2 \alpha_{\rm in}^2}
\end{equation}

From Eq.~(\ref{daoutQNL}) we have
\begin{eqnarray}
\delta \tilde Y_{\rm out} (\omega)  &=&   - 2 \eta  (g_{\rm sig}' + \Delta')   \delta \tilde X_{\rm in} +  \left [ 1 - 2 \eta \left (1 - i  \omega'   \right )  \right ] \delta \tilde Y_{\rm in} -  2\sqrt{\eta (1-\eta)} \left [  (g_{\rm sig}' + \Delta') \delta X_l  + \left (1 - i \omega'  \right ) \delta Y_l \right ]\\
&& - 4 \alpha_{\rm in} \left (g_{\rm sig}' + \Delta' \right ) \left [1  - 2 \eta - i \eta \omega' \right ]  \delta \gamma_{\rm in}'  - 4 \eta  \alpha_{\rm in}   \left (1 - i \omega'  \right ) {g_{\rm therm}^{\rm self}}' 
\end{eqnarray}
where we have used the relations $\delta \gamma_{\rm in}' (-\omega)^\dagger = \delta \gamma_{\rm in}' (\omega)$ and $ {g_{\rm therm}^{\rm self}}' (-\omega)^\dagger =  {g_{\rm therm}^{\rm self}}' (\omega)$ which arise since both the input coupling rate and thermal fluctuations are real parameters in the time domain; and in general $\tilde X_{\rm out}(\omega) =  \tilde a^\dagger (-\omega) + \tilde a(\omega) $, and $\tilde Y(\omega) = i \left [ \tilde a^\dagger(-\omega) - \tilde a (\omega) \right ]$. Given that each of the fluctuating terms in this expression are uncorrelated, the output phase quadrature variance is
\begin{eqnarray}
\left \langle \left | \delta \tilde Y_{\rm out} (\omega) \right |^2 \right \rangle  &=&   4 \eta^2  (g_{\rm sig}' + \Delta')^2   \left \langle \delta \tilde X_{\rm in}^2 \right \rangle +  \left [ ( 1 - 2 \eta)^2 + 4 \eta^2 {\omega'}^2    \right ] \left \langle \delta \tilde Y_{\rm in}^2 \right \rangle +  4 \eta (1-\eta) \left [  (g_{\rm sig}' + \Delta')^2 \left \langle \delta X_l^2 \right \rangle  + \left (1 + {\omega'}^2  \right ) \left \langle \delta Y_l^2 \right \rangle \right ]\\
&& + 16 \alpha_{\rm in}^2 \left (g_{\rm sig}' + \Delta' \right )^2 \left [(1  - 2 \eta)^2 + \eta^2 {\omega'}^2 \right ] \left \langle {\delta \gamma_{\rm in}'}^2 \right \rangle  + 16 \eta^2  \alpha_{\rm in}^2   \left (1 + {\omega'}^2  \right ) \left \langle {{g_{\rm therm}^{\rm self}}'}^2 \right \rangle 
\end{eqnarray} 

The laser input variances can be broken into a quantum part and a classical part, as described by Eqs.~(\ref{VXin})~and~(\ref{VYin}); whilst the fields entering through the sensor loss channel are vacuum and therefore quantum noise limited ($ \langle \delta X_l^2 \rangle=\langle \delta Y_l^2  \rangle=1$). Using these relations we find
\begin{eqnarray}
\left \langle \left | \delta \tilde Y_{\rm out} (\omega) \right |^2 \right \rangle  &=&  4 \eta^2  (g_{\rm sig}' + \Delta')^2   \left (1 + V_{\rm RIN} \alpha_{\rm in}^2 \right ) +  \left [ ( 1 - 2 \eta)^2 + 4 \eta^2 {\omega'}^2    \right ] \left ( 1 + V_\zeta \alpha_{\rm in}^2 \right ) \\
&& +  4 \eta (1-\eta) \left [  (g_{\rm sig}' + \Delta')^2 + \left (1 + {\omega'}^2  \right ) \right ]\\
&& + 16 \alpha_{\rm in}^2 \left (g_{\rm sig}' + \Delta' \right )^2 \left [(1  - 2 \eta)^2 + \eta^2 {\omega'}^2 \right ] \left \langle {\delta \gamma_{\rm in}'}^2 \right \rangle  + 16 \eta^2  \alpha_{\rm in}^2   \left (1 + {\omega'}^2  \right ) \left \langle {{g_{\rm therm}^{\rm self}}'}^2 \right \rangle \\
&=&  4 \eta^2  (g_{\rm sig}' + \Delta')^2   V_{\rm RIN} \alpha_{\rm in}^2  +  \left [ ( 1 - 2 \eta)^2 + 4 \eta^2 {\omega'}^2    \right ]  V_\zeta \alpha_{\rm in}^2  \\
&& + 16 \alpha_{\rm in}^2 \left (g_{\rm sig}' + \Delta' \right )^2 \left [(1  - 2 \eta)^2 + \eta^2 {\omega'}^2 \right ] \left \langle {\delta \gamma_{\rm in}'}^2 \right \rangle  + 16 \eta^2  \alpha_{\rm in}^2   \left (1 + {\omega'}^2  \right ) \left \langle {{g_{\rm therm}^{\rm self}}'}^2 \right \rangle \\
&& + 1 + 4 \eta \left [ ( g_{\rm sig}' + \Delta')^2 +{\omega'}^2 \right ]
\end{eqnarray}
Taking the reasonable limit that $\{ {g_{\rm sig}}', \Delta' , \omega' \} \ll 1$, and that the presence of the particle to be sensed does not effect the noise (i.e. neglecting all noise terms multiplied by ${g_{\rm sig}}'$), we have
\begin{equation}
\left \langle \left | \delta \tilde Y_{\rm out} (\omega) \right |^2 \right \rangle  =  4 \eta^2 \alpha_{\rm in}^2  {\Delta'}^2   V_{\rm RIN}   +  \alpha_{\rm in}^2 \left ( 1 - 2 \eta \right )^2  V_\zeta    + 16 \alpha_{\rm in}^2  {\Delta'}^2 \left (1  - 2 \eta \right )^2  \left \langle {\delta \gamma_{\rm in}'}^2 \right \rangle  + 16 \eta^2  \alpha_{\rm in}^2  \left \langle {{g_{\rm therm}^{\rm self}}'}^2 \right \rangle +1. \label{VY_QNL}
\end{equation}
This expression shows the effect of each of the four sources of noise on variance of the output field. Notice that since  $ \Delta'  \ll 1$, the relative intensity noise $V_{\rm RIN}$ is greatly suppressed compared to the laser phase noise $V_\zeta$ in this measurement.

The uncertainty in our estimate of $g_{\rm sig}'$ is then finally
\begin{eqnarray}
\left \langle \left | {\delta g_{\rm sig}^{\rm est}}' \right |^2 \right \rangle_{\rm QNL} &=&  \left ( \frac{\Delta'}{2}\right )^2   V_{\rm RIN}   +   \left ( \frac{ 1 - 2 \eta}{4 \eta} \right )^2  V_\zeta    + {\Delta'}^2 \left (\frac{1  - 2 \eta}{\eta} \right )^2  \left \langle {\delta \gamma_{\rm in}'}^2 \right \rangle  + \left \langle {{g_{\rm therm}^{\rm self}}'}^2 \right \rangle +\frac{1}{16 \eta^2 \alpha_{\rm in}^2}\\
&=&\left ( \frac{\Delta'}{2}\right )^2   V_{\rm RIN}   +   \left ( \frac{ 1 - 2 \eta}{4 \eta} \right )^2  V_\zeta    + {\Delta'}^2 \left (\frac{1  - 2 \eta}{\eta} \right )^2 V_\gamma + V_{\rm therm} +\frac{1}{16 \eta^2 \alpha_{\rm in}^2},
 \label{QNLsense}
\end{eqnarray}
where for compactness we have defined $\left \langle {\delta \gamma_{\rm in}'}^2 \right \rangle = V_\gamma$ and $ \left \langle {{g_{\rm therm}^{\rm self}}'}^2 \right \rangle = V_{\rm therm}$. From this expression one sees that, as expected, only the contribution of shot noise towards the uncertainty (the final term) depends on laser power. Of the other terms, it should be expected that laser phase noise dominates, since this is typically large in the frequency range of interest to biosensing experiments and is not suppressed by the measurement technique in the same way that intensity noise is.

\section{Backscatter measurement}

To perform a similar calculation to that performed above for backscatter measurement, we return to the full Hamiltonian in Eq.~(\ref{H}). The equations of motion for the two intracavity annihilation operators are then
\begin{eqnarray}
\dot{\hat a}_A &=& \frac{1}{i \hbar} \left [\hat a_A, H \right ] - \gamma (t) \hat a_A + \sqrt{2 \gamma_{\rm in}(t)} \hat a_{A,\rm in} + \sqrt{2 \gamma_l} \delta \hat a_{A,l}\\
&=& i \left ( g_{\rm sig} + g^{\rm self}_{\rm therm} + \Delta \right ) \hat a_A + i \left (g_{\rm sig} e^{2 i m \phi_{\rm sig}} + g_0e^{2 i m \phi_0} + g_{\rm therm}^{\rm cross} \right ) \hat a_C - \gamma (t) \hat a_A  + \sqrt{2 \gamma_{\rm in}(t)} \hat a_{A,\rm in} + \sqrt{2 \gamma_l} \delta \hat a_{A,l} \label{eqnmoA} \\
\dot{\hat a}_C &=& \frac{1}{i \hbar} \left [\hat a_C, H \right ] - \gamma (t) \hat a_C + \sqrt{2 \gamma_{\rm in}(t)} \delta \hat a_{C,\rm in} + \sqrt{2 \gamma_l} \delta \hat a_{C,l}\\
&=& i \left ( g_{\rm sig} + g^{\rm self}_{\rm therm} + \Delta \right ) \hat a_C + i \left (g_{\rm sig} e^{-2 i m \phi_{\rm sig}} + g_0e^{-2 i m \phi_0} + {g_{\rm therm}^{\rm cross}}^* \right ) \hat a_A - \gamma (t) \hat a_C  + \sqrt{2 \gamma_{\rm in}(t)} \delta \hat a_{C,\rm in} + \sqrt{2 \gamma_l} \delta \hat a_{C,l} \label{eqnmoC}
\end{eqnarray}

\subsection{Solving the mean fields}

Similar to the above, we can find the steady state mean fields by taking the expectation values of Eqs.~(\ref{eqnmoA})~and~(\ref{eqnmoC}), setting the time derivative to zero, and solving. From Eq.~(\ref{eqnmoA}) we find
\begin{equation}
\alpha_A \left  [  \bar \gamma - i \left ( g_{\rm sig}  + \Delta \right )  \right ] = i \left ( g_{\rm sig} e^{2 i m \phi_{\rm sig}} + g_0e^{2 i m \phi_0}  \right ) \alpha_C + \sqrt{2 \bar \gamma_{\rm in}} \alpha_{A,\rm in}
\end{equation}
Rearranging we find
\begin{equation}
\alpha_A  = \frac{ i \left ( g_{\rm sig} e^{2 i m \phi_{\rm sig}} + g_0e^{2 i m \phi_0}  \right ) \alpha_C + \sqrt{2 \bar \gamma_{\rm in}} \alpha_{A,\rm in}}{ \bar \gamma - i \left ( g_{\rm sig}  + \Delta \right ) }.
\end{equation}
Similarly for $\alpha_C$ we have
\begin{equation}
\alpha_C = \frac{ i \left ( g_{\rm sig} e^{-2 i m \phi_{\rm sig}} + g_0e^{-2 i m \phi_0}  \right ) }{ \bar \gamma - i \left ( g_{\rm sig}  + \Delta \right ) } \alpha_A,
\end{equation}
where here, we consider a scenario where only the anti-clockwise mode is directly pumped, and therefore $\alpha_{C,\rm in}=0$. It is then straightforward to solve for $\alpha_C$ as
\begin{eqnarray}
\alpha_C &=& \left [ \frac{ i \left ( g_{\rm sig} e^{-2 i m \phi_{\rm sig}} + g_0e^{-2 i m \phi_0}  \right )}{ \bar \gamma - i \left ( g_{\rm sig}  + \Delta \right ) }  \right ] \left [ \frac{ i \left ( g_{\rm sig} e^{2 i m \phi_{\rm sig}} + g_0e^{2 i m \phi_0}  \right ) \alpha_C + \sqrt{2 \bar \gamma_{\rm in}} \alpha_{A,\rm in}}{ \bar \gamma - i \left ( g_{\rm sig}  + \Delta \right ) } \right ],\\
&& \hspace{-10mm} \left [ 1 +  \frac{\left | g_{\rm sig} e^{2 i m \phi_{\rm sig}} + g_0e^{2 i m \phi_0}  \right |^2 }{\left ( \bar \gamma - i \left ( g_{\rm sig}  + \Delta \right ) \right )^2} \right ] \alpha_C = \left [ \frac{i \sqrt{2 \bar \gamma_{\rm in}}  \left ( g_{\rm sig} e^{-2 i m \phi_{\rm sig}} + g_0e^{-2 i m \phi_0}  \right )}{\left ( \bar \gamma - i \left ( g_{\rm sig}  + \Delta \right ) \right )^2} \right ] \alpha_{A, \rm in}\\
\alpha_C &=& \left [ \frac{i \sqrt{2 \bar \gamma_{\rm in}}  \left ( g_{\rm sig} e^{-2 i m \phi_{\rm sig}} + g_0e^{-2 i m \phi_0}  \right )}{\left ( \bar \gamma - i \left ( g_{\rm sig}  + \Delta \right ) \right )^2 + \left | g_{\rm sig} e^{2 i m \phi_{\rm sig}} + g_0e^{2 i m \phi_0}  \right |^2} \right ] \alpha_{A, \rm in}\\
&=& \left [ \frac{i \sqrt{2 \bar \gamma_{\rm in}}  \left ( g_{\rm sig} e^{-2 i m \phi_{\rm sig}} + g_0e^{-2 i m \phi_0}  \right )}{\left ( \bar \gamma - i \Delta \right )^2} \right ] \alpha_{A, \rm in}\\
&=& i \frac{ \sqrt{2 \bar \gamma_{\rm in}}}{\left ( \bar \gamma - i \Delta \right )^2}  \left ( g_{\rm sig} e^{-2 i m \phi_{\rm sig}} + g_0 e^{-2 i m \phi_0}  \right ) \alpha_{A, \rm in}\\
\end{eqnarray}
where we've used the assumption that $\{ g_{\rm sig}, g_0 \} \ll \bar \gamma$. We can then find the intracavity field amplitude in mind $A$
\begin{eqnarray}
\alpha_A &=& \left [ \frac{{-\sqrt{2 \bar \gamma_{\rm in}}}  \left | g_{\rm sig} e^{2 i m \phi_{\rm sig}} + g_0 e^{2 i m \phi_0}  \right |^2/ \left(\bar \gamma - i \Delta \right )^2 + \sqrt{2 \bar \gamma_{\rm in}} }{ \bar \gamma - i \left ( g_{\rm sig}  + \Delta \right ) } \right ] \alpha_{A, \rm in}\\
&=&  \sqrt{2 \bar \gamma_{\rm in}} \left [ \frac{\left(\bar \gamma - i \Delta \right )^2  - \left | g_{\rm sig} e^{2 i m \phi_{\rm sig}} + g_0 e^{2 i m \phi_0}  \right |^2  }{{\left ( \bar \gamma - i \left ( g_{\rm sig}  + \Delta \right ) \right ) \left(\bar \gamma - i \Delta \right )^2}  } \right ] \alpha_{A, \rm in}\\
&\approx& \frac{\sqrt{2 \bar \gamma_{\rm in}}}{\left ( \bar \gamma - i \Delta \right )} \left [ 1 - \frac{ g_0^2 }{{ \left(\bar \gamma - i \Delta \right )^2}  } \right ] \alpha_{A, \rm in},
\end{eqnarray}
Where we have, as usual, assumed that $ g_{\rm sig} \ll \{ g_0, \bar \gamma \}$. Furthermore, since $ g_0 \ll \bar \gamma$ we immediately see that $\alpha_A \gg \alpha_C$. One might, therefore think that a higher signal-to-noise could be obtained by detecting $\alpha_A$ that $\alpha_C$. However, this is not this case. In fact, when the incident field is on resonance with the cavity resonance any signal arising from the field output from mode $A$ is second order in $g_{\rm sig}$, whereas that arising from the field output from mode $C$ is first order.

Since $\alpha_{C, \rm out} = \alpha_{C,\rm in} - \sqrt{2 \bar \gamma_{\rm in}} \alpha_C$, and $\alpha_{C,\rm in} =0$ we have
\begin{equation}
\alpha_{C,\rm out} = -  \frac{ 2 i \eta}{\left (1 - i \Delta' \right )^2}  \left ( g_{\rm sig}' e^{-2 i m \phi_{\rm sig}} + g_0' e^{-2 i m \phi_0}  \right ) \alpha_{A, \rm in} \label{alphaCout}
\end{equation}

\subsection{Solving for the fluctuating noise}

Taking the fluctuating parts of Eqs.~(\ref{eqnmoA})~and~(\ref{eqnmoC}) we have
\begin{eqnarray}
\delta \dot{\hat a}_A \!\!\!
&=& \!\!\! - \left [  \bar \gamma \!- \!i \left ( g_{\rm sig} \!+\! \Delta \right ) \right ]\delta \hat a_A + i \left (g_{\rm sig} e^{2 i m \phi_{\rm sig}} \!+\! g_0e^{2 i m \phi_0}  \right ) \delta \hat a_C  + i  \alpha_A g^{\rm self}_{\rm therm}     + i \alpha_C g_{\rm therm}^{\rm cross}  + \left [\frac{\alpha_{A, \rm in}}{\sqrt{2 \bar \gamma_{\rm in}}} \! -\! \alpha_A  \right ] \delta \gamma_{\rm in}   + \! \sqrt{2 \bar \gamma_{\rm in}} \delta a_{A,\rm in} + \! \sqrt{2 \gamma_l} \delta \hat a_{A,l}  \nonumber \\
\delta \dot{\hat a}_C \!\!\!
&=& \!\!\! - \left [  \bar \gamma \!- \!i \left ( g_{\rm sig} \!+ \!\Delta \right ) \right ]\delta \hat a_C + i \left (g_{\rm sig} e^{-2 i m \phi_{\rm sig}} \!+\! g_0e^{-2 i m \phi_0}  \right ) \delta \hat a_A  + i  \alpha_C g^{\rm self}_{\rm therm}     + i \alpha_A{ g_{\rm therm}^{\rm cross}}^* \!\! - \!\alpha_C  \delta \gamma_{\rm in}   + \! \sqrt{2 \bar \gamma_{\rm in}} \delta a_{C,\rm in} + \! \sqrt{2 \gamma_l} \delta \hat a_{C,l} \nonumber
\end{eqnarray}

Taking the Fourier transform and rearranging we find
\begin{eqnarray}
\delta \tilde a_A \! \! 
&=& \! \! \frac{1}{ \bar \gamma + i \left ( \omega - g_{\rm sig} - \Delta \right )} \left \{  i \left (g_{\rm sig} e^{2 i m \phi_{\rm sig}} \!+\! g_0e^{2 i m \phi_0}  \right ) \delta \tilde a_C  + i  \alpha_A g^{\rm self}_{\rm therm}     + i \alpha_C g_{\rm therm}^{\rm cross}  + \left [\frac{\alpha_{A, \rm in}}{\sqrt{2 \bar \gamma_{\rm in}}} \! -\! \alpha_A  \right ] \delta \gamma_{\rm in}   + \! \sqrt{2 \bar \gamma_{\rm in}} \delta \tilde a_{A,\rm in} + \! \sqrt{2 \gamma_l} \delta \tilde a_{A,l} \right \}
 \nonumber \\
\delta \tilde a_C \! \!
&=& \! \! \frac{1}{\bar \gamma + i \left ( \omega - g_{\rm sig} - \Delta \right )} \left \{  i \left (g_{\rm sig} e^{-2 i m \phi_{\rm sig}} \!+\! g_0e^{-2 i m \phi_0}  \right ) \delta \tilde a_A  + i  \alpha_C g^{\rm self}_{\rm therm}     + i \alpha_A{ g_{\rm therm}^{\rm cross}}^* \!\!  - \!\alpha_C \delta \gamma_{\rm in}   + \! \sqrt{2 \bar \gamma_{\rm in}} \delta \tilde a_{C,\rm in} + \! \sqrt{2 \gamma_l} \delta \tilde a_{C,l} \right \} \nonumber
\end{eqnarray}
Substituting the first of these equations into the second, we can find an expression for $\delta \hat a_C$ in terms of the input noise sources.
\begin{eqnarray}
 \left [ \bar \gamma + i \left ( \omega - g_{\rm sig} - \Delta \right ) \right ] \delta \tilde a_C \! \!
&=&     \frac{i \left (g_{\rm sig} e^{-2 i m \phi_{\rm sig}} \!+\! g_0e^{-2 i m \phi_0}  \right )}{ \bar \gamma + i \left ( \omega - g_{\rm sig} - \Delta \right )} \Bigg \{  i \left (g_{\rm sig} e^{2 i m \phi_{\rm sig}} \!+\! g_0e^{2 i m \phi_0}  \right ) \delta \tilde a_C  + i  \alpha_A g^{\rm self}_{\rm therm}     + i \alpha_C g_{\rm therm}^{\rm cross}  + \left [\frac{\alpha_{A, \rm in}}{\sqrt{2 \bar \gamma_{\rm in}}} \! -\! \alpha_A  \right ] \delta \gamma_{\rm in} \nonumber  \\
&& + \! \sqrt{2 \bar \gamma_{\rm in}} \delta \tilde a_{A,\rm in} + \! \sqrt{2 \gamma_l} \delta \tilde a_{A,l} \Bigg \} + i  \alpha_C g^{\rm self}_{\rm therm}     + i \alpha_A{ g_{\rm therm}^{\rm cross}}^* \!\! - \!\alpha_C  \delta \gamma_{\rm in}   + \! \sqrt{2 \bar \gamma_{\rm in}} \delta \tilde a_{C,\rm in} + \! \sqrt{2 \gamma_l} \delta \tilde a_{C,l} \nonumber\\
&& \hspace{-40mm} \left \{ \left [ \bar \gamma + i \left ( \omega - g_{\rm sig} - \Delta \right ) \right ]^2  +  \left | g_{\rm sig} e^{2 i m \phi_{\rm sig}} \!+\! g_0e^{2 i m \phi_0} \right |^2\right \} \delta \tilde a_C \!  = \\
&&\!  i \left (g_{\rm sig} e^{-2 i m \phi_{\rm sig}} \!+\! g_0e^{-2 i m \phi_0}  \right ) \Bigg \{  i  \alpha_A g^{\rm self}_{\rm therm}     + i \alpha_C g_{\rm therm}^{\rm cross}  + \left [\frac{\alpha_{A, \rm in}}{\sqrt{2 \bar \gamma_{\rm in}}} \! -\! \alpha_A  \right ] \delta \gamma_{\rm in} \nonumber + \! \sqrt{2 \bar \gamma_{\rm in}} \delta \tilde a_{A,\rm in} + \! \sqrt{2 \gamma_l} \delta \tilde a_{A,l} \Bigg \} \\
&&+ \left [  \bar \gamma + i \left ( \omega - g_{\rm sig} - \Delta \right ) \right ] \left \{ i  \alpha_C g^{\rm self}_{\rm therm}     + i \alpha_A{ g_{\rm therm}^{\rm cross}}^* \!\!  - \!\alpha_C   \delta \gamma_{\rm in}   + \! \sqrt{2 \bar \gamma_{\rm in}} \delta \tilde a_{C,\rm in} + \! \sqrt{2 \gamma_l} \delta \tilde a_{C,l} \right \}\nonumber\\
 \left [ \bar \gamma + i \left ( \omega - g_{\rm sig} - \Delta \right ) \right ]^2  \delta \tilde a_C \! 
& \approx & \! \left \{  i \alpha_C  \left [  \bar \gamma + i \left ( \omega - g_{\rm sig} - \Delta \right ) \right ]  - \alpha_A g_0e^{-2 i m \phi_0}    \right \} g^{\rm self}_{\rm therm}  + i \alpha_A  \left [  \bar \gamma + i \left ( \omega - g_{\rm sig} - \Delta \right ) \right ]  { g_{\rm therm}^{\rm cross}}^* - \alpha_C  g_0e^{-2 i m \phi_0} g_{\rm therm}^{\rm cross} \nonumber\\
 && - \left \{    \alpha_C \left [  \bar \gamma + i \left ( \omega - g_{\rm sig} - \Delta \right ) \right ]   -   i  g_0e^{-2 i m \phi_0}  \left [\frac{\alpha_{A, \rm in}}{\sqrt{2 \bar \gamma_{\rm in}}} \! -\! \alpha_A  \right ]  \right \} \delta \gamma_{\rm in}\\
 && +  i g_0e^{-2 i m \phi_0}  \left (   \! \sqrt{2 \bar \gamma_{\rm in}} \delta \tilde a_{A,\rm in} + \! \sqrt{2 \gamma_l} \delta \tilde a_{A,l} \right ) \nonumber   + \left [  \bar \gamma + i \left ( \omega - g_{\rm sig} - \Delta \right ) \right ]  \left ( \sqrt{2 \bar \gamma_{\rm in}} \delta \tilde a_{C,\rm in} + \! \sqrt{2 \gamma_l} \delta \tilde a_{C,l} \right ) \nonumber\\
\end{eqnarray}

Substituting $g_{\rm therm}^{\rm cross}  = R_{\rm therm}^{\rm cross}  + i I_{\rm therm}^{\rm cross}$, we have
\begin{eqnarray}
 \left [ \bar \gamma + i \left ( \omega - g_{\rm sig} - \Delta \right ) \right ]^2  \delta \tilde a_C \! 
& =& \! \left \{  i \alpha_C  \left [  \bar \gamma + i \left ( \omega - g_{\rm sig} - \Delta \right ) \right ]  - \alpha_A g_0e^{-2 i m \phi_0}    \right \} g^{\rm self}_{\rm therm}  \\
&&  + \left \{  i \alpha_A  \left [  \bar \gamma + i \left ( \omega - g_{\rm sig} - \Delta \right ) \right ] - \alpha_C  g_0e^{-2 i m \phi_0} \right \} R_{\rm therm}^{\rm cross} \\
&& + \left \{ \alpha_A  \left [  \bar \gamma + i \left ( \omega - g_{\rm sig} - \Delta \right ) \right ]   - i \alpha_C  g_0e^{-2 i m \phi_0} \right \} I_{\rm therm}^{\rm cross}\\
 && - \left \{  \alpha_C  \left [  \bar \gamma + i \left ( \omega - g_{\rm sig} - \Delta \right ) \right ]     -  i  g_0e^{-2 i m \phi_0}  \left [\frac{\alpha_{A, \rm in}}{\sqrt{2 \bar \gamma_{\rm in}}} \! -\! \alpha_A  \right ]  \right \} \delta \gamma_{\rm in}\\
 && +  i g_0e^{-2 i m \phi_0}  \left (   \! \sqrt{2 \bar \gamma_{\rm in}} \delta \tilde sa_{A,\rm in} + \! \sqrt{2 \gamma_l} \delta \tilde a_{A,l} \right ) \nonumber   + \left [  \bar \gamma + i \left ( \omega - g_{\rm sig} - \Delta \right ) \right ]  \left ( \sqrt{2 \bar \gamma_{\rm in}} \tilde a_{C,\rm in} + \! \sqrt{2 \gamma_l} \delta \tilde a_{C,l} \right ) \nonumber\\
 & =& \! \left \{  i \alpha_C  \left [  \bar \gamma + i \left ( \omega - g_{\rm sig} - \Delta \right ) \right ]  - \alpha_A g_0e^{-2 i m \phi_0}    \right \} g^{\rm self}_{\rm therm}  \\
&&  +  i \alpha_A  \left [  \bar \gamma + i \left ( \omega - g_{\rm sig} - \Delta \right ) \right ]   R_{\rm therm}^{\rm cross} + \alpha_A  \left [  \bar \gamma + i \left ( \omega - g_{\rm sig} - \Delta \right ) \right ]   I_{\rm therm}^{\rm cross}\\
 && - \left \{    \alpha_C \left [  \bar \gamma + i \left ( \omega - g_{\rm sig} - \Delta \right ) \right ]      -   i  g_0e^{-2 i m \phi_0}  \left [\frac{\alpha_{A, \rm in}}{\sqrt{2 \bar \gamma_{\rm in}}} \! -\! \alpha_A  \right ]  \right \} \delta \gamma_{\rm in}\\
 && +  i g_0e^{-2 i m \phi_0}  \left (   \! \sqrt{2 \bar \gamma_{\rm in}} \delta \tilde a_{A,\rm in} + \! \sqrt{2 \gamma_l} \delta \tilde a_{A,l} \right ) \nonumber   + \left [  \bar \gamma + i \left ( \omega - g_{\rm sig} - \Delta \right ) \right ]  \left ( \sqrt{2 \bar \gamma_{\rm in}} \delta \tilde a_{C,\rm in} + \! \sqrt{2 \gamma_l} \delta \tilde a_{C,l} \right )
\end{eqnarray}
where we have neglected product terms of $g_0$ and $\alpha_C$ since $g_0 \ll \bar \gamma$ and $\alpha_C \ll \alpha_A$.

The output field fluctuations $\delta \tilde a_{C,out}$ can be related to the input through $\delta \hat a_{C,\rm out} = \delta \tilde a_{C, \rm in} - \sqrt{2 \bar \gamma_{\rm in}} \delta \tilde a_C - \alpha_C/\sqrt{2 \bar \gamma_{\rm in}} \delta \gamma_{\rm in}$. Defining 
\begin{equation}
\delta \tilde a_{C,\rm out} = A {g^{\rm self}_{\rm therm}}' + B {R_{\rm therm}^{\rm cross}}'  + C {I_{\rm therm}^{\rm cross}}' + D \delta \gamma_{\rm in}' + E  \delta \tilde a_{A,\rm in}  + F  \delta \tilde a_{A,l}  + G \delta \tilde a_{C,\rm in}  + H \delta \tilde a_{C,l},
\end{equation}
we then have
\begin{eqnarray}
A&=& - \sqrt{2 \bar \gamma_{\rm in}} \bar \gamma \frac{  \left \{ i \alpha_C  \left [  \bar \gamma + i \left ( \omega - g_{\rm sig} - \Delta \right ) \right ]  - \alpha_A g_0e^{-2 i m \phi_0} \right \}}{ \left [ \bar \gamma + i \left ( \omega - g_{\rm sig} - \Delta \right ) \right ]^2 } \\
B &=&  - \frac {i \bar \gamma \alpha_A \sqrt{2 \bar \gamma_{\rm in}}} { \bar \gamma + i \left ( \omega - g_{\rm sig} - \Delta \right )}\\
C &=&  - \frac {\bar \gamma \alpha_A \sqrt{2 \bar \gamma_{\rm in}}} { \bar \gamma + i \left ( \omega - g_{\rm sig} - \Delta \right )}\\
D &=&   \bar \gamma  \frac{   \! \sqrt{2 \bar \gamma_{\rm in}} \alpha_C \left [  \bar \gamma + i \left ( \omega - g_{\rm sig} - \Delta \right ) \right ] -   i  g_0e^{-2 i m \phi_0}  \left ( \alpha_{A, \rm in}\! -\! \sqrt{2 \bar \gamma_{\rm in}} \alpha_A  \right )  }{\left [ \bar \gamma + i \left ( \omega - g_{\rm sig} - \Delta \right ) \right ]^2 }  - \frac{\bar \gamma \alpha_C}{\sqrt{2 \bar \gamma_{\rm in}}}\\
E  &=& - \frac{2  i \bar \gamma_{\rm in} g_0e^{-2 i m \phi_0}  }{\left [ \bar \gamma + i \left ( \omega - g_{\rm sig} - \Delta \right ) \right ]^2 } \approx - \frac{2  i \eta g_0'e^{-2 i m \phi_0}  }{\left [ 1 + i( \omega'-  \Delta') \right ]^2 } \\
F &=& - \frac{2  i \sqrt{ \bar \gamma_{\rm in} \gamma_i } g_0e^{-2 i m \phi_0}  }{\left [ \bar \gamma + i \left ( \omega - g_{\rm sig} - \Delta \right ) \right ]^2 } \approx    - \frac{2  i \sqrt{ \eta (1 - \eta) } g_0'e^{-2 i m \phi_0}  }{\left ( 1 - i \Delta' \right )^2 }   \\
G &=& 1 -  \frac{2 \bar \gamma_{\rm in} }{ \bar \gamma + i \left ( \omega - g_{\rm sig} - \Delta \right )} \approx  1 -  \frac{2 \eta }{1 - i \Delta'}\\
H &=& -  \frac{2 \sqrt{ \bar \gamma_{\rm in} \gamma_l} }{ \bar \gamma + i \left ( \omega - g_{\rm sig} - \Delta \right )} \approx -  \frac{2 \sqrt{\eta (1-\eta) } }{ 1- i  \Delta' },
\end{eqnarray}
where in the expressions for $E$-$H$ we have moved into dimensionless units, and made the approximation that $\{ \omega, g_{\rm sig} \} \ll \bar \gamma$. In $E$ we retain the $\omega'$ dependence since, latter in the calculation the other terms will cancel, leaving it as the dominant term. To simplify $A$-$D$ we must substitute in for $\alpha_A$ and $\alpha_C$.
\begin{eqnarray}
A&=& \frac{2 \bar \gamma_{\rm in} \bar \gamma g_0 e^{-2 i m \phi_0}}{\left ( \bar \gamma - i \Delta \right )^2}  \frac{  \left \{  \left [  \bar \gamma + i \left ( \omega - g_{\rm sig} - \Delta \right ) \right ]  + \left [ \bar \gamma - i \Delta - \frac{ g_0^2 }{{ \left(\bar \gamma - i \Delta \right )}  } \right ] \right \}}{ \left [ \bar \gamma + i \left ( \omega - g_{\rm sig} - \Delta \right ) \right ]^2 } \alpha_{A, \rm in} \\
&\approx&  {2 \bar \gamma_{\rm in} \bar \gamma g_0 e^{-2 i m \phi_0}} \left ( \frac{  2 \bar \gamma + i \left ( \omega - 2 \Delta \right )   }{\left ( \bar \gamma - i \Delta \right )^2  \left [ \bar \gamma + i \left ( \omega  - \Delta \right ) \right ]^2 } \right ) \alpha_{A, \rm in} \\
&=& {2 \eta g_0' e^{-2 i m \phi_0}} \left ( \frac{  2 + i \left ( \omega' - 2 \Delta' \right )   }{\left ( 1- i \Delta' \right )^2  \left [ 1 + i \left ( \omega'  - \Delta' \right ) \right ]^2 } \right ) \alpha_{A, \rm in} \\
&\approx& \frac{4 \eta g_0' e^{-2 i m \phi_0}} {\left ( 1- i \Delta' \right )^3 }  \alpha_{A, \rm in},
\end{eqnarray}
taking $\{ g_{\rm sig}, \omega \} \ll \bar \gamma$, and $g_0^2 \ll \bar \gamma^2$.

\begin{eqnarray}
B &=& -   \frac{2 i \bar \gamma \bar \gamma_{\rm in}}{\left ( \bar \gamma - i \Delta \right )^2} \left [ 1 - \frac{ g_0^2 }{{ \left(\bar \gamma - i \Delta \right )^2}  } \right ] \alpha_{A, \rm in}\\
&=& -   {2 i \eta} \left ( \frac{ { \left(1 - i \Delta' \right )^2}  -  {g_0'}^2 }{{\left ( 1 - i \Delta' \right )^4}} \right ) \alpha_{A, \rm in}
\end{eqnarray}

\begin{eqnarray}
C &=&  -iB
\end{eqnarray}

\begin{eqnarray}
D &=&   \frac{ i \bar \gamma   g_0 e^{-2 i m \phi_0} }{\left ( \bar \gamma - i \Delta \right )^2}   \left ( \frac{    2 \bar \gamma_{\rm in}    \alpha_{A, \rm in} }{\bar \gamma - i   \Delta }    -   2 \alpha_{A, \rm in} \! +\! \sqrt{2 \bar \gamma_{\rm in}} \alpha_A   \right )\\
&=&  \frac{ i \bar \gamma   g_0 e^{-2 i m \phi_0} }{\left ( \bar \gamma - i \Delta \right )^3}   \left ( 2 \bar \gamma_{\rm in}    \alpha_{A, \rm in}      -   2 \alpha_{A, \rm in} \left ( \bar \gamma - i   \Delta \right ) \! +\! 2 \bar \gamma_{\rm in} \left [ 1 - \frac{ g_0^2 }{{ \left(\bar \gamma - i \Delta \right )^2}  } \right ] \alpha_{A, \rm in}  \right )\\
&=&  \frac{ 2 i \bar \gamma   g_0 e^{-2 i m \phi_0} }{\left ( \bar \gamma - i \Delta \right )^3}   \left [ 2 \bar \gamma_{\rm in}        -   \bar \gamma + i   \Delta  \right ] \alpha_{A, \rm in}\\
&=& \frac{ 2 i g_0' e^{-2 i m \phi_0} }{\left (1 - i \Delta' \right )^3} \left [ 2 \eta - 1 + i   \Delta'  \right ] \alpha_{A, \rm in}
\end{eqnarray}

The output field can now be determined as 
\begin{eqnarray}
\tilde a_{C,\rm out} &=& \alpha_{C, \rm out} + \delta \tilde a_{C,\rm out}\\
&=&  -  \frac{ 2 i \eta}{\left (1 - i \Delta' \right )^2}  \left ( g_{\rm sig}' e^{-2 i m \phi_{\rm sig}} + g_0' e^{-2 i m \phi_0}  \right ) \alpha_{A, \rm in} + \delta \tilde a_{C,\rm out}
\end{eqnarray}

Let us determine the signal that is received on direct detection of this field, i.e.
\begin{eqnarray}
i &=& \tilde a_{C,\rm out}^\dagger (- \omega) \tilde a_{C,\rm out} (\omega)\\
&=& \frac{ 4 \eta^2}{\left (1 + {\Delta'}^2 \right )^2}  \left | g_{\rm sig}' e^{2 i m \phi_{\rm sig}} + g_0' e^{2 i m \phi_0}  \right |^2 \alpha_{A, \rm in}^2 -  \frac{ 2 i \eta}{\left (1 - i \Delta' \right )^2}   g_0' e^{-2 i m \phi_0}  \alpha_{A, \rm in}  \delta \tilde a_{C,\rm out}^\dagger +  \frac{ 2 i \eta}{\left (1 + i \Delta' \right )^2}  g_0' e^{2 i m \phi_0}  \alpha_{A, \rm in}  \delta \tilde a_{C,\rm out}\\
&=& \frac{ 4 g_{0}' \eta^2 \alpha_{A, \rm in}^2}{\left (1 + {\Delta'}^2 \right )^2}  \left [ g_{0}' + g_{\rm sig}'  \left ( e^{- 2 i m ( \phi_{\rm sig} - \phi_0)}  + e^{2 i m ( \phi_{\rm sig} - \phi_0)} \right ) \right ]  +  2 i \eta g_0'  \alpha_{A, \rm in} \left (\frac{e^{2 i m \phi_0}}{\left (1 + i \Delta' \right )^2}  \delta \tilde a_{C,\rm out} - \frac{e^{-2 i m \phi_0}}{\left (1 - i \Delta' \right )^2}  \delta \tilde a_{C,\rm out}^\dagger \right )\\
&=& \frac{ 4 g_{0}' \eta^2 \alpha_{A, \rm in}^2}{\left (1 + {\Delta'}^2 \right )^2}  \left [ g_{0}' + 2 g_{\rm sig}' \cos 2 m ( \phi_{\rm sig} - \phi_0)  \right ]  +  2 i \eta g_0'  \alpha_{A, \rm in} \left (\frac{e^{2 i m \phi_0}}{\left (1 + i \Delta' \right )^2}  \delta \tilde a_{C,\rm out} - \frac{e^{-2 i m \phi_0}}{\left (1 - i \Delta' \right )^2}  \delta \tilde a_{C,\rm out}^\dagger \right )
\end{eqnarray}
neglecting the noise product term $\delta \tilde a_{C,\rm out}^\dagger (- \omega) \delta \tilde a_{C,\rm out} (\omega)$, and, since the signal is small the product terms of signal $g_{\rm sig}'$ and noise $\delta \tilde a_{C,\rm out}$, and the signal product term ${g_{\rm sig}'}^2$.

The mean measured signal is then
\begin{equation}
\langle i \rangle = \frac{ 4 g_{0}' \eta^2 \alpha_{A, \rm in}^2}{\left (1 + {\Delta'}^2 \right )^2}  \left [ g_{0}' + 2 g_{\rm sig}' \cos 2 m ( \phi_{\rm sig} - \phi_0)  \right ] 
\end{equation}
Rearranging this expression in terms of $g_{\rm sig}'$ we have
\begin{equation}
g_{\rm sig}' = \frac{1}{2 \cos 2 m ( \phi_{\rm sig} - \phi_0) } \left [ \frac{\left (1 + {\Delta'}^2 \right )^2}{4 g_{0}' \eta^2 \alpha_{A, \rm in}^2} \langle i \rangle - g_0' \right ],
\end{equation}
such that, based on the measurement $i$ an estimate of  $g_{\rm sig}'$ may be formed as
\begin{equation}
{g_{\rm sig}^{\rm est}}' =  g_{\rm sig}' = \frac{1}{2 \cos 2 m ( \phi_{\rm sig} - \phi_0) } \left [ \frac{\left (1 + {\Delta'}^2 \right )^2}{4 g_{0}' \eta^2 \alpha_{A, \rm in}^2} i  - g_0' \right ].
\end{equation}
The uncertainty in this estimate is
\begin{eqnarray}
\left \langle  \left | \delta {g_{\rm sig}^{\rm est}}' \right |^2 \right \rangle_{\rm back-scatter} &=&  \frac{\left (1 + {\Delta'}^2 \right )^4}{64 {g_{0}'}^2 \eta^4 \alpha_{A, \rm in}^4 \cos^2 2 m ( \phi_{\rm sig} - \phi_0) } \left \langle \left | \delta i \right |^2 \right \rangle  \\
&\approx& \frac{ \left \langle \left | \delta i \right |^2 \right \rangle}{64 {g_{0}'}^2 \eta^4 \alpha_{A, \rm in}^4 \cos^2 2 m ( \phi_{\rm sig} - \phi_0) } \label{Vgest}
\end{eqnarray}
Now we must find $ \left \langle \left | \delta i \right |^2 \right \rangle$
\begin{eqnarray}
\left \langle \left | \delta i \right |^2 \right \rangle &=&  4 \eta^2 {g_0'}^2  \alpha_{A, \rm in}^2 \left \langle \left | \frac{e^{2 i m \phi_0}}{\left (1 + i \Delta' \right )^2}  \delta \tilde a_{C,\rm out}(\omega) - \frac{e^{-2 i m \phi_0}}{\left (1 - i \Delta' \right )^2}  \delta \tilde a_{C,\rm out}^\dagger(-\omega)  \right |^2 \right \rangle \\
&=& \frac{4 \eta^2 {g_0'}^2  \alpha_{A, \rm in}^2}{\left ( 1 + {\Delta'}^2 \right )^2} \left \langle \left | e^{2 i m \breve{\phi}}  \delta \tilde a_{C,\rm out}(\omega)  - e^{-2 i m \breve{\phi}} \delta \tilde a_{C,\rm out}^\dagger (-\omega)  \right |^2 \right \rangle, \label{Vi}
\end{eqnarray}
where for conciseness we have combined the phase shift due to detuning with that due to $g_0'$, making the definitions $(1 + {\Delta'}^2)/(1 + i \Delta')^{2} = e^{2 i m \phi_\Delta}$, and $\breve{\phi} = \phi_0 + \phi_\Delta$.
In general the expectation value may be expanded in terms of a series of noise operators $\delta \tilde a_j$ each with coefficients $A_j$ (in our case these coefficients are $A$ through $H$, and the noise operators are those associated to each of $A$ through $H$).
\begin{eqnarray}
\left \langle \left | e^{2 i m \breve{\phi}}  \delta \tilde a_{C,\rm out}(\omega)  - e^{-2 i m \breve{\phi}} \delta \tilde a_{C,\rm out}^\dagger (- \omega) \right |^2 \right \rangle &=& \left \langle \left | e^{2 i m \breve{\phi}} \sum_jA_j(\omega)  \delta \tilde a_{j}(\omega)  - e^{-2 i m \breve{\phi}} \sum_j  A^*_j(-\omega)  \delta \tilde a_{j}^\dagger(-\omega)  \right |^2 \right \rangle\\
&=& \sum_j  \left \langle \left | e^{2 i m \breve{\phi}}  A_j(\omega)  \delta \tilde a_{j}(\omega)  - e^{-2 i m \breve{\phi}}  A^*_j(-\omega)  \delta \tilde a_{j}^\dagger(-\omega)  \right |^2 \right \rangle \label{sumeq}
\end{eqnarray}
where we have assumed, as is the relevant case here, that all noise sources are uncorrelated. It therefore only remains to calculate the contributions to the variance from each noise source.

\begin{eqnarray}
\left \langle \left | e^{2 i m \breve{\phi}} A {g_{\rm therm}^{\rm self}}' - e^{-2 i m \breve{\phi}}  A^* {{g_{\rm therm}^{\rm self}}'}^* \right |^2 \right \rangle &=&    \left | e^{2 i m \breve{\phi}} A - e^{-2 i m \breve{\phi}}  A^*  \right |^2 \left \langle  {{g_{\rm therm}^{\rm self}}'} ^2 \right \rangle \\
&=&  \left | e^{2 i m \breve{\phi}} \frac{4 \eta g_0' e^{-2 i m \phi_0}} {\left ( 1- i \Delta' \right )^3 }  \alpha_{A, \rm in} - e^{-2 i m \breve{\phi}}  \frac{4 \eta g_0' e^{2 i m \phi_0}} {\left ( 1+ i \Delta' \right )^3 }  \alpha_{A, \rm in}  \right |^2 \left \langle {{g_{\rm therm}^{\rm self}}'} ^2 \right \rangle\\
&=& 16 \eta^2 {g_0'}^2   \alpha_{A, \rm in}^2 \left | \frac{1 + \Delta'^2} {\left ( 1+ \Delta'^2 \right ) ^2 \left ( 1- i \Delta' \right ) }   -  \frac{1 + \Delta'^2} {\left ( 1+ \Delta'^2 \right ) ^2 \left ( 1+ i \Delta' \right ) }   \right |^2 \left \langle {{g_{\rm therm}^{\rm self}}'} ^2 \right \rangle\\
&=& \frac{16 \eta^2 {g_0'}^2   \alpha_{A, \rm in}^2}{\left ( 1+ \Delta'^2 \right ) ^2} \left | \frac{1}{ 1- i \Delta' }   -  \frac{1 } { 1+ i \Delta' }   \right |^2 \left \langle {{g_{\rm therm}^{\rm self}}'} ^2 \right \rangle\\
&=& 64 \eta^2 {g_0'}^2   \alpha_{A, \rm in}^2 \frac{\Delta'^2}{\left ( 1+ \Delta'^2 \right ) ^4}  \left \langle {{g_{\rm therm}^{\rm self}}'} ^2 \right \rangle
\end{eqnarray}
where we've used the fact that $e^{2im \breve{\phi}}e^{- 2im\phi_0}=e^{2im\phi_\Delta} = (1 + {\Delta'}^2)/(1+i \Delta')^2$.
From this, we see that, in contrast to sensing using a transmitted phase shift, when using backscattered light, the self-thermorefractive noise is greatly suppressed. If the cavity is put exactly on resonance $\Delta=0$, then self-thermorefractive noise is exactly eliminated.

\begin{eqnarray}
\left \langle \left | e^{2 i m \breve{\phi}} B {R_{\rm therm}^{\rm cross}}' - e^{-2 i m \breve{\phi}}  B^* {{R_{\rm therm}^{\rm cross}}'}^* \right |^2 \right \rangle &=&     \left | e^{2 i m \breve{\phi}} B - e^{-2 i m \breve{\phi}}  B^*  \right |^2  \left \langle  {{R_{\rm therm}^{\rm cross}}'}^2 \right \rangle\\
&=&   \left | - e^{2 i m \breve{\phi}}  {2 i \eta} \left ( \frac{ { \left(1 - i \Delta' \right )^2}  -  {g_0'}^2 }{{\left ( 1 - i \Delta' \right )^4}} \right ) \alpha_{A, \rm in} - e^{-2 i m \breve{\phi}}  {2 i \eta} \left ( \frac{ { \left(1 + i \Delta' \right )^2}  -  {g_0'}^2 }{{\left ( 1 + i \Delta' \right )^4}} \right ) \alpha_{A, \rm in} \right |^2  \left \langle {{R_{\rm therm}^{\rm cross}}'} ^2 \right \rangle \nonumber\\
&=& 4 \eta^2  \alpha_{A, \rm in}^2  \left | e^{2 i m \breve{\phi}}   \left ( \frac{ { \left(1 - i \Delta' \right )^2}  -  {g_0'}^2 }{{\left ( 1 - i \Delta' \right )^4}} \right )  + e^{-2 i m \breve{\phi}}   \left ( \frac{ { \left(1 + i \Delta' \right )^2}  -  {g_0'}^2 }{{\left ( 1 + i \Delta' \right )^4}} \right ) \right |^2   \left \langle {{R_{\rm therm}^{\rm cross}}'} ^2 \right \rangle \nonumber\\
&=& 4 \eta^2  \alpha_{A, \rm in}^2   \left | e^{2 i m \breve{\phi}}   \left ( \frac{ (1 + {\Delta'}^2) e^{2 i m \phi_\Delta} }{{\left ( 1 + {\Delta'}^2 \right )^2 e^{4im\phi_\Delta}}} \right )  + e^{-2 i m \breve{\phi}}   \left ( \frac{ {(1 + {\Delta'}^2) e^{-2i m \phi_\Delta}} }{\left ( 1 + {\Delta'}^2 \right )^2 e^{-4im\phi_\Delta} } \right ) \right |^2  \left \langle  {{R_{\rm therm}^{\rm cross}}'}^2 \right \rangle \nonumber\\
&=& \frac{4 \eta^2  \alpha_{A, \rm in}^2}{\left ( 1 + {\Delta'}^2 \right )^2}   \left | e^{2 i m \phi_0}  + e^{-2 i m \phi_0}  \right |^2  \left \langle {{R_{\rm therm}^{\rm cross}}'} ^2 \right \rangle \nonumber\\
&=& \frac{16 \eta^2  \alpha_{A, \rm in}^2}{\left ( 1 + {\Delta'}^2 \right )^2}  \cos^2 {2 m \phi_0}   \left \langle {{R_{\rm therm}^{\rm cross}}'} ^2 \right \rangle \label{Rcrossexp}\\
&\approx& 16 \eta^2  \alpha_{A, \rm in}^2 \cos^2 2 m  \phi_0  \left \langle {{R_{\rm therm}^{\rm cross}}'} ^2 \right \rangle, 
\end{eqnarray}
where we have taken $g_0' \ll 1$. From this we see that the cross-thermorefractive noise does contribute to the measured noise in the back-scatter detection protocol, so it is not possible to entirely avoid thermo-refractive noise.

\begin{eqnarray}
\left \langle \left | e^{2 i m \breve{\phi}} C {I_{\rm therm}^{\rm cross}}' - e^{-2 i m \breve{\phi}}  C^* {{I_{\rm therm}^{\rm cross}}'}^* \right |^2 \right \rangle &=&     \left | e^{2 i m \breve{\phi}} C - e^{-2 i m \breve{\phi}}  C^*  \right |^2  \left \langle {{I_{\rm therm}^{\rm cross}}'} ^2 \right \rangle\\
&=&  \left | i e^{2 i m \breve{\phi}} B + i e^{-2 i m \breve{\phi}}  B^*  \right |^2  \left \langle  {{I_{\rm therm}^{\rm cross}}'} ^2 \right \rangle\\
&=&  \left | e^{2 i m \breve{\phi}} B + e^{-2 i m \breve{\phi}}  B^*  \right |^2  \left \langle {{R_{\rm therm}^{\rm cross}}'} ^2 \right \rangle\\
&=& \frac{4 \eta^2  \alpha_{A, \rm in}^2}{\left ( 1 + {\Delta'}^2 \right )^2}   \left | e^{2 i m \phi_0}  - e^{-2 i m \phi_0}  \right |^2  \left \langle {{R_{\rm therm}^{\rm cross}}'} ^2 \right \rangle \nonumber\\
&=& \frac{16 \eta^2  \alpha_{A, \rm in}^2}{\left ( 1 + {\Delta'}^2 \right )^2}  \sin^2 {2 m \phi_0}   \left \langle {{R_{\rm therm}^{\rm cross}}'} ^2 \right \rangle \label{Icrossexp}\\
&\approx& 16 \eta^2  \alpha_{A, \rm in}^2 \sin^2 2 m  \phi_0  \left \langle {{R_{\rm therm}^{\rm cross}}'} ^2 \right \rangle, 
\end{eqnarray}
where we've used the fact that $\left \langle  {{I_{\rm therm}^{\rm cross}}'}^2 \right \rangle = \left \langle  {{R_{\rm therm}^{\rm cross}}'}^2 \right \rangle$ (Eq.~(\ref{therm_relation})). 

From Eqs.~(\ref{Rcrossexp})~and~(\ref{Icrossexp}) it can be seen that the total contribution of thermorefractive noise to the measured photocurrent is
\begin{eqnarray}
\left \langle \left | e^{2 i m \breve{\phi}} B {R_{\rm therm}^{\rm cross}}' - e^{-2 i m \breve{\phi}}  B^* {{R_{\rm therm}^{\rm cross}}'}^* \right |^2 \right \rangle +\left \langle \left | e^{2 i m \breve{\phi}} C {I_{\rm therm}^{\rm cross}}' - e^{-2 i m \breve{\phi}}  C^* {{I_{\rm therm}^{\rm cross}}'}^* \right |^2 \right \rangle &=& \frac{16 \eta^2  \alpha_{A, \rm in}^2}{\left ( 1 + {\Delta'}^2 \right )^2}   \left \langle {{R_{\rm therm}^{\rm cross}}'} ^2 \right \rangle \label{Rcrossexp}\\
&\approx& 16 \eta^2  \alpha_{A, \rm in}^2  \left \langle {{R_{\rm therm}^{\rm cross}}'} ^2 \right \rangle.
\end{eqnarray}

\begin{eqnarray}
\left \langle \left | e^{2 i m \breve{\phi}} D \delta \gamma_{\rm in}' - e^{-2 i m \breve{\phi}}  D^* { \delta \gamma_{\rm in}' }^* \right |^2 \right \rangle &=&     \left | e^{2 i m \breve{\phi}} D - e^{-2 i m \breve{\phi}}  D^*  \right |^2  \left \langle  { \delta \gamma_{\rm in}' }^2 \right \rangle\\
&=& \left | e^{2 i m \breve{\phi}} \frac{ 2 i g_0' e^{-2 i m \phi_0} }{\left (1 - i \Delta' \right )^3} \left [ 2 \eta - 1 + i   \Delta'  \right ] \alpha_{A, \rm in} + e^{-2 i m \breve{\phi}}  \frac{ 2 i g_0' e^{2 i m \phi_0} }{\left (1 + i \Delta' \right )^3} \left [ 2 \eta - 1 - i   \Delta'  \right ] \alpha_{A, \rm in}  \right |^2  \left \langle {\delta \gamma_{\rm in}'  }^2 \right \rangle \nonumber\\
&=& 4 {g_0'}^2 \alpha_{A, \rm in}^2 \left |  \frac{ e^{2 i m \phi_\Delta} }{\left (1 - i \Delta' \right )^3} \left [ 2 \eta - 1 + i   \Delta'  \right ] +  \frac{ e^{-2 i m \phi_\Delta} }{\left (1 + i \Delta' \right )^3} \left [ 2 \eta - 1 - i   \Delta'  \right ]  \right |^2  \left \langle { \delta \gamma_{\rm in}'  }^2 \right \rangle \nonumber\\
&=& 4 {g_0'}^2 \alpha_{A, \rm in}^2 \left |  \frac{ 1 + {\Delta'}^2 }{\left (1 - i \Delta' \right )^3 (1 + i \Delta')^2} \left [ 2 \eta - 1 + i   \Delta'  \right ] +  \frac{1 + {\Delta'}^2}{\left (1 + i \Delta' \right )^3 (1 - i \Delta')^2}  \left [ 2 \eta - 1 - i   \Delta'  \right ]  \right |^2  \left \langle {  \delta \gamma_{\rm in}'  }^2 \right \rangle \nonumber\\
&=& \frac{4 {g_0'}^2 \alpha_{A, \rm in}^2}{\left ( 1 + \Delta'^2 \right )^2} \left |  \frac{ 2 \eta - 1 + i   \Delta'   }{\left (1 - i \Delta' \right )} +  \frac{ 2 \eta - 1 - i   \Delta' }{\left (1 + i \Delta' \right )}   \right |^2  \left \langle { \delta \gamma_{\rm in}'  }^2 \right \rangle \nonumber\\
&=&  \frac{4 {g_0'}^2 \alpha_{A, \rm in}^2}{\left ( 1 + \Delta'^2 \right )^2}  \left |  \frac{ 2 \eta - 1 + 2 i \eta  \Delta'  -{\Delta'}^2  }{1 + \Delta'^2} + \frac{ 2 \eta - 1 - 2 i \eta  \Delta'  - {\Delta'}^2  }{1 + \Delta'^2}   \right |^2  \left \langle { \delta \gamma_{\rm in}'  }^2 \right \rangle \nonumber\\
&=&16 {g_0'}^2 \alpha_{A, \rm in}^2 \left [ \frac{ 2 \eta - 1   -{\Delta'}^2 }{\left ( 1 + \Delta'^2 \right )^2  }\right ]^2  \left \langle { \delta \gamma_{\rm in}'  }^2 \right \rangle \nonumber\\
&\approx& 16 {g_0'}^2 \alpha_{A, \rm in}^2 \left ( 2 \eta - 1  \right )^2  \left \langle {  \delta \gamma_{\rm in}'  }^2 \right \rangle \nonumber
\end{eqnarray}
This expression is identical to the noise term due to input coupling fluctuations in the direct measurement case 
(see Eq.~\ref{VY_QNL}), with the exception that here $\Delta' = \Delta_0' + g_0'$ is replaced with simply $g_0'$. Hence, a disadvantage of the backscatter scheme is that input coupling noise cannot be removed by simply ensuring the light in on resonance with the cavity. However, it should be noticed, that this noise source may be arbitrarily suppressed in principle by sitting close to critical coupling where $\eta=0.5$.

\begin{eqnarray}
\left \langle \left | e^{2 i m \breve{\phi}} E(\omega) \delta \tilde a_{A,\rm in}- e^{-2 i m \breve{\phi}}  E^*(-\omega) { \delta \tilde a_{A,\rm in} }^\dagger \right |^2 \right \rangle &=&    \frac{1}{4}\left \langle \left | e^{2 i m \breve{\phi}} E \left (\delta \tilde X_{A,\rm in} + i \delta \tilde Y_{A,\rm in} \right )- e^{-2 i m \breve{\phi}}  E^* \left (\delta \tilde X_{A,\rm in} - i \delta \tilde Y_{A,\rm in} \right ) \right |^2 \right \rangle\\
&=& \frac{1}{4}\left \langle \left | e^{2 i m \breve{\phi}} E \delta \tilde X_{A,\rm in} - e^{-2 i m \breve{\phi}}  E^* \delta \tilde X_{A,\rm in} \right |^2 \right \rangle 
+ \frac{1}{4}\left \langle \left | e^{2 i m \breve{\phi}} E \delta \tilde Y_{A,\rm in} + e^{-2 i m \breve{\phi}}  E^* \delta \tilde Y_{A,\rm in}  \right |^2 \right \rangle \nonumber \\
&=&\frac{1}{4}\left | e^{2 i m \breve{\phi}} E(\omega) - e^{-2 i m \breve{\phi}}  E^* (-\omega) \right |^2 \! \! \left \langle  \delta \tilde X_{A,\rm in}^2 \right \rangle 
+ \frac{1}{4} \left | e^{2 i m \breve{\phi}} E(\omega)  + e^{-2 i m \breve{\phi}}  E^*(-\omega)   \right |^2  \! \! \left \langle \delta \tilde Y_{A,\rm in}^2 \right \rangle. \nonumber
\end{eqnarray}
\begin{eqnarray}
\left | e^{2 i m \breve{\phi}} E(\omega) - e^{-2 i m \breve{\phi}}  E^* (-\omega) \right |^2 &=& \left | e^{2 i m \breve{\phi}}  \frac{2  i \eta g_0'e^{-2 i m \phi_0}  }{\left [1 + i \left (\omega' - \Delta' \right ) \right ]^2 }  + e^{-2 i m \breve{\phi}}   \frac{2  i \eta g_0'e^{2 i m \phi_0}  }{\left [1 + i \left (\omega' + \Delta' \right ) \right ]^2} \right |^2\\
&=&  4  \eta^2 { g_0'}^2 \left (1 + \Delta'^2 \right )^2 \left |  \frac{1}{\left ( 1 + i  \Delta' \right )^2 \left [1 + i \left (\omega' - \Delta' \right ) \right ]^2    }  + \frac{1 }{\left ( 1 - i  \Delta' \right )^2    \left [1 + i \left (\omega' + \Delta' \right ) \right ]^2  } \right |^2\\
&=&  4  \eta^2 { g_0'}^2 \left (1 + \Delta'^2 \right )^2 \left |  \frac{\left ( 1 - i  \Delta' \right )^2    \left [1 + i \left (\omega' + \Delta' \right ) \right ]^2   +   \left ( 1 + i  \Delta' \right )^2 \left [1 + i \left (\omega' - \Delta' \right ) \right ]^2    }{\left ( 1 + \Delta'^2 \right )^2 \left [1 + i \left (\omega' - \Delta' \right ) \right ]^2   \left [1 + i \left (\omega' + \Delta' \right ) \right ]^2  } \right |^2\\
&=&  4  \eta^2 { g_0'}^2 \frac{ \left |  \left ( 1\! - \! 2 i  \Delta' \! - \! \Delta'^2\right ) \!  \left [1\! +\! 2 i \left (\omega' \! + \! \Delta' \right ) \! - \!  \left ( \omega' \! + \! \Delta'\right )^2\right ]  \!  +   \! \left ( 1 \! + \! 2 i  \Delta' \! - \! \Delta'^2\right )\! \left [1 \! + \! 2 i \left (\omega' \! - \! \Delta' \right ) \! - \! \left (\omega' \! - \! \Delta' \right )^2 \right ]  \right |^2  }{  \left (1 + \Delta'^2 \right )^2 \left [1 + \left (\omega' - \Delta' \right )^2 \right ]^2   \left [1 +  \left (\omega' + \Delta' \right )^2 \right ]^2  } \nonumber \\
&=&  16  \eta^2 { g_0'}^2 \frac{ \left ( 1 - \omega'^2 + 2 \Delta'^2 + \Delta'^2 \omega'^2 + \Delta'^4  \right )^2 + 4 \omega'^2 \left ( 1 + \Delta'^2\right )^2  }{  \left (1 + \Delta'^2 \right )^2 \left [1 + \left (\omega' - \Delta' \right )^2 \right ]^2   \left [1 +  \left (\omega' + \Delta' \right )^2 \right ]^2  } \\
&\approx&  16  \eta^2 { g_0'}^2 \frac{ \left ( 1 + 2 \Delta'^2  + \Delta'^4  \right )^2   }{  \left (1 + \Delta'^2 \right )^2 \left (1 +  \Delta'^2 \right )^2   \left ( 1 +  \Delta'^2 \right )^2  } \label{approx1} \\
&=&   \frac{  16 \eta^2 { g_0'}^2 }{  \left (1 + \Delta'^2 \right )^2  } \\                                                     
&\approx& 16 \eta^2 { g_0'}^2, \label{approx2}
\end{eqnarray}
where at Eq.~(\ref{approx1}) we have made the approximation $\omega' \ll 1$, and at Eq.~(\ref{approx2}) we have made the approximation $\Delta' \ll 1$. Similarly, for the second term in $E$ we find
\begin{eqnarray}
\left | e^{2 i m \breve{\phi}} E(\omega) + e^{-2 i m \breve{\phi}}  E^* (-\omega) \right |^2 &=& \left | e^{2 i m \breve{\phi}}  \frac{2  i \eta g_0'e^{-2 i m \phi_0}  }{\left [1 + i \left (\omega' - \Delta' \right ) \right ]^2 }  - e^{-2 i m \breve{\phi}}   \frac{2  i \eta g_0'e^{2 i m \phi_0}  }{\left [1 + i \left (\omega' + \Delta' \right ) \right ]^2} \right |^2\\
&=&  4  \eta^2 { g_0'}^2 \left (1 + \Delta'^2 \right )^2 \left |  \frac{1}{\left ( 1 + i  \Delta' \right )^2 \left [1 + i \left (\omega' - \Delta' \right ) \right ]^2    }  - \frac{1 }{\left ( 1 - i  \Delta' \right )^2    \left [1 + i \left (\omega' + \Delta' \right ) \right ]^2  } \right |^2\\
&=&  4  \eta^2 { g_0'}^2 \left (1 + \Delta'^2 \right )^2 \left |  \frac{\left ( 1 - i  \Delta' \right )^2    \left [1 + i \left (\omega' + \Delta' \right ) \right ]^2   -   \left ( 1 + i  \Delta' \right )^2 \left [1 + i \left (\omega' - \Delta' \right ) \right ]^2    }{\left ( 1 + \Delta'^2 \right )^2 \left [1 + i \left (\omega' - \Delta' \right ) \right ]^2   \left [1 + i \left (\omega' + \Delta' \right ) \right ]^2  } \right |^2\\
&=&  4  \eta^2 { g_0'}^2 \frac{ \left |  \left ( 1\! - \! 2 i  \Delta' \! - \! \Delta'^2\right ) \!  \left [1\! +\! 2 i \left (\omega' \! + \! \Delta' \right ) \! - \!  \left ( \omega' \! + \! \Delta'\right )^2\right ]  \!  -   \! \left ( 1 \! + \! 2 i  \Delta' \! - \! \Delta'^2\right )\! \left [1 \! + \! 2 i \left (\omega' \! - \! \Delta' \right ) \! - \! \left (\omega' \! - \! \Delta' \right )^2 \right ]  \right |^2  }{  \left (1 + \Delta'^2 \right )^2 \left [1 + \left (\omega' - \Delta' \right )^2 \right ]^2   \left [1 +  \left (\omega' + \Delta' \right )^2 \right ]^2  } \nonumber \\
&=&  64  \eta^2 { g_0'}^2 \frac{ \omega'^2 \Delta'^2 \left ( 1 + \omega'^2 \right )^2 + \Delta'^2 \omega'^4   }{  \left (1 + \Delta'^2 \right )^2 \left [1 + \left (\omega' - \Delta' \right )^2 \right ]^2   \left [1 +  \left (\omega' + \Delta' \right )^2 \right ]^2  } \\
&=&  64  \eta^2 { g_0'}^2 \omega'^2 \Delta'^2 \frac{  \left ( 1 + \omega'^2 \right )^2 + \omega'^2   }{  \left (1 + \Delta'^2 \right )^2 \left [1 + \left (\omega' - \Delta' \right )^2 \right ]^2   \left [1 +  \left (\omega' + \Delta' \right )^2 \right ]^2  }  \\
&\approx&   \frac{64  \eta^2 { g_0'}^2 \omega'^2 \Delta'^2 }{  \left (1 + \Delta'^2 \right )^6 }  \\
&\approx& 64  \eta^2 { g_0'}^2 \omega'^2 \Delta'^2.
\end{eqnarray}
We therefore find
\begin{eqnarray}
\left \langle \left | e^{2 i m \breve{\phi}} E \delta \tilde a_{A,\rm in}- e^{-2 i m \breve{\phi}}  E^* { \delta \tilde a_{A,\rm in} }^\dagger \right |^2 \right \rangle &=& \frac{  4 \eta^2 { g_0'}^2 }{  \left (1 + \Delta'^2 \right )^2  }  \left \langle  \delta \tilde X_{A,\rm in}^2 \right \rangle +   \frac{16  \eta^2 { g_0'}^2 \omega'^2 \Delta'^2 }{  \left (1 + \Delta'^2 \right )^6 } \left \langle \delta \tilde Y_{A,\rm in}^2 \right \rangle  \label{Eeq}\\
&\approx& 4   \eta^2 { g_0'}^2  \left \langle  \delta \tilde X_{A,\rm in}^2 \right \rangle + 16 \eta^2 { g_0'}^2 \omega'^2 \Delta'^2  \left \langle \delta \tilde Y_{A,\rm in}^2 \right \rangle
\end{eqnarray}

Substituting for the amplitude and phase noise of the incident field from Eqs.~(\ref{VXin})~and~(\ref{VYin}) we have
\begin{eqnarray}
\left \langle \left | e^{2 i m \breve{\phi}} E \delta \tilde a_{A,\rm in}- e^{-2 i m \breve{\phi}}  E^* { \delta \tilde a_{A,\rm in} }^\dagger \right |^2 \right \rangle &=& 
 \frac{  4 \eta^2 { g_0'}^2 }{  \left (1 + \Delta'^2 \right )^2  } \left ( 1 + V_{\rm RIN} (\omega) \alpha_{A, \rm in}^2  \right )  +   \frac{16  \eta^2 { g_0'}^2 \omega'^2 \Delta'^2 }{  \left (1 + \Delta'^2 \right )^6 } \left (1 + V_{\zeta} (\omega) \alpha_{A, \rm in}^2 \right ) \\
 &\approx& \frac{ 4  \eta^2 { g_0'}^2}{ \left (1 + \Delta'^2 \right )^2} \left [ 1 +  V_{\rm RIN} (\omega) \alpha_{A, \rm in}^2 + \frac{4 \omega'^2 \Delta'^2 }{  \left (1 + \Delta'^2 \right )^4 } V_{\zeta} (\omega) \alpha_{A, \rm in}^2 \right ]\\
  &\approx&  4  \eta^2 { g_0'}^2 \left [ 1 +  \alpha_{A,in}^2 \left ( V_{\rm RIN} (\omega) + 4 \omega'^2 \Delta'^2 V_{\zeta} (\omega) \right ) \right ]
\end{eqnarray}
We see that there is a fundamental noise floor due to shot noise (the $1$), whilst laser phase noise is suppressed relative to intensity noise by a factor given by $4 \omega'^2 {\Delta'}^2$.

\begin{eqnarray}
\left \langle \left | e^{2 i m \breve{\phi}} F \delta \tilde a_{A,l}- e^{-2 i m \breve{\phi}}  F^* { \delta \tilde a_{A,l} }^\dagger \right |^2 \right \rangle &=&    \frac{1}{4}\left \langle \left | e^{2 i m \breve{\phi}} F \left (\delta \tilde X_{A,l} + i \delta \tilde Y_{A,l} \right )- e^{-2 i m \breve{\phi}}  F^* \left (\delta \tilde X_{A,l} - i \delta \tilde Y_{A,l} \right ) \right |^2 \right \rangle\\
&=& \frac{1}{4}\left \langle \left | e^{2 i m \breve{\phi}} F \delta \tilde X_{A,l} - e^{-2 i m \breve{\phi}}  F^* \delta \tilde X_{A,l} \right |^2 \right \rangle 
+ \frac{1}{4}\left \langle \left | e^{2 i m \breve{\phi}} F \delta \tilde Y_{A,l} + e^{-2 i m \breve{\phi}}  F^* \delta \tilde Y_{A,l}  \right |^2 \right \rangle\\
&=&\frac{1}{4}\left | e^{2 i m \breve{\phi}} F - e^{-2 i m \breve{\phi}}  F^*\right |^2  \left \langle  \delta \tilde X_{A,l}^2 \right \rangle 
+ \frac{1}{4} \left | e^{2 i m \breve{\phi}} F  + e^{-2 i m \breve{\phi}}  F^*   \right |^2  \left \langle \delta \tilde Y_{A,l}^2 \right \rangle\\
&=& \frac{  4 \eta(1-\eta) { g_0'}^2 }{  \left (1 + \Delta'^2 \right )^2  }  \left \langle  \delta \tilde X_{A,l}^2 \right \rangle +   \frac{16  \eta(1-\eta) { g_0'}^2 \omega'^2 \Delta'^2 }{  \left (1 + \Delta'^2 \right )^6 } \left \langle \delta \tilde Y_{A,l}^2 \right \rangle  \label{mideq}\\
&=& \frac{  4 \eta(1-\eta) { g_0'}^2 }{  \left (1 + \Delta'^2 \right )^2  } +   \frac{16  \eta(1-\eta) { g_0'}^2 \omega'^2 \Delta'^2 }{  \left (1 + \Delta'^2 \right )^6 } \label{vacstep}\\
&=& \frac{  4 \eta(1-\eta) { g_0'}^2 }{  \left (1 + \Delta'^2 \right )^2  } \left [ 1 +   \frac{4  \omega'^2 \Delta'^2 }{  \left (1 + \Delta'^2 \right )^4 } \right ]\\
&\approx& 4 \eta(1-\eta) { g_0'}^2,
\end{eqnarray}
where to get to Eq.~(\ref{mideq}) we have used Eq.~(\ref{Eeq}) combined with the fact that $F$ is identical to $E$ except for the substitution $\eta^2 \rightarrow \eta(1-\eta)$, and to get to Eq.~(\ref{vacstep}) we have used the fact that the incident field entering the system through the cavity loss channels is in a vacuum state, and therefore $\left \langle \delta \tilde X_{A,l}^2 \right \rangle=\left \langle \delta \tilde Y_{A,l}^2 \right \rangle=1$.

\begin{eqnarray}
\left \langle \left | e^{2 i m \breve{\phi}} G \delta \tilde a_{C,\rm in}- e^{-2 i m \breve{\phi}}  G^* { \delta \tilde a_{C,\rm in} }^\dagger \right |^2 \right \rangle &=&    \frac{1}{4}\left \langle \left | e^{2 i m \breve{\phi}} G \left (\delta \tilde X_{C,\rm in} + i \delta \tilde Y_{C,\rm in} \right )- e^{-2 i m \breve{\phi}}  G^* \left (\delta \tilde X_{C,\rm in} - i \delta \tilde Y_{C,\rm in} \right ) \right |^2 \right \rangle\\
&=& \frac{1}{4}\left \langle \left | e^{2 i m \breve{\phi}} G \delta \tilde X_{C,\rm in} - e^{-2 i m \breve{\phi}}  G^* \delta \tilde X_{C,\rm in} \right |^2 \right \rangle 
+ \frac{1}{4}\left \langle \left | e^{2 i m \breve{\phi}} G \delta \tilde Y_{C,\rm in} + e^{-2 i m \breve{\phi}}  G^* \delta \tilde Y_{C,\rm in}  \right |^2 \right \rangle\\
&=&\frac{1}{4}\left | e^{2 i m \breve{\phi}} G - e^{-2 i m \breve{\phi}}  G^*\right |^2  \left \langle  \delta \tilde X_{C,\rm in}^2 \right \rangle 
+ \frac{1}{4} \left | e^{2 i m \breve{\phi}} G  + e^{-2 i m \breve{\phi}}  G^*   \right |^2  \left \langle \delta \tilde Y_{C,\rm in}^2 \right \rangle\\
&=& \frac{1}{4}\left | e^{2 i m \breve{\phi}} G - e^{-2 i m \breve{\phi}}  G^*\right |^2  + \frac{1}{4} \left | e^{2 i m \breve{\phi}} G  + e^{-2 i m \breve{\phi}}  G^*   \right |^2 \label{vacstep2} \\
&=& \frac{1}{4}\left ( 2 |G|^2 - 2 |G|^2 + 2 |G|^2 + 2 |G|^2   \right )\\
&=& |G|^2\\
&=& \left | 1 -  \frac{2 \eta }{1 - i \Delta'} \right |^2\\
&=& \frac{(1- 2 \eta)^2 + {\Delta'}^2}{1 + {\Delta'}^2}\\
&\approx& (1- 2 \eta)^2+ {\Delta'}^2
\end{eqnarray}
where to arrive at Eq.~(\ref{vacstep2}) we have used the fact that the incident field entering the system through the input channel into mode $C$ is in a vacuum state, and therefore $\left \langle \delta \tilde X_{C,\rm in}^2 \right \rangle=\left \langle \delta \tilde Y_{C, \rm in}^2 \right \rangle=1$.

\begin{eqnarray}
\left \langle \left | e^{2 i m \breve{\phi}} H \delta \tilde a_{C,l}- e^{-2 i m \breve{\phi}}  H^* { \delta \tilde a_{C,l} }^\dagger \right |^2 \right \rangle &=&    \frac{1}{4}\left \langle \left | e^{2 i m \breve{\phi}} H \left (\delta \tilde X_{C,l} + i \delta \tilde Y_{C,l} \right )- e^{-2 i m \breve{\phi}}  H^* \left (\delta \tilde X_{C,l} - i \delta \tilde Y_{C,l} \right ) \right |^2 \right \rangle\\
&=& \frac{1}{4}\left \langle \left | e^{2 i m \breve{\phi}} H \delta \tilde X_{C,l} - e^{-2 i m \breve{\phi}}  H^* \delta \tilde X_{C,l} \right |^2 \right \rangle 
+ \frac{1}{4}\left \langle \left | e^{2 i m \breve{\phi}} H \delta \tilde Y_{C,l} + e^{-2 i m \breve{\phi}}  H^* \delta \tilde Y_{C,l}  \right |^2 \right \rangle\\
&=&\frac{1}{4}\left | e^{2 i m \breve{\phi}} H - e^{-2 i m \breve{\phi}}  H^*\right |^2  \left \langle  \delta \tilde X_{C,l}^2 \right \rangle 
+ \frac{1}{4} \left | e^{2 i m \breve{\phi}} H  + e^{-2 i m \breve{\phi}}  H^*   \right |^2  \left \langle \delta \tilde Y_{C,l}^2 \right \rangle\\
&=& \frac{1}{4}\left | e^{2 i m \breve{\phi}} H - e^{-2 i m \breve{\phi}}  H^*\right |^2  + \frac{1}{4} \left | e^{2 i m \breve{\phi}} H  + e^{-2 i m \breve{\phi}}  H^*   \right |^2 \label{vacstep3} \\
&=& \frac{1}{4}\left ( 2 |H|^2 - 2 |H|^2 + 2 |H|^2 + 2 |H|^2   \right )\\
&=& |H|^2\\
&=& \left |   -  \frac{2 \sqrt{\eta (1-\eta) } }{ 1- i  \Delta' } \right |^2\\
&=& \frac{4 \eta (1-\eta)}{1 + {\Delta'}^2}\\
&\approx& 4 \eta (1-\eta)
\end{eqnarray}
where to arrive at Eq.~(\ref{vacstep3}) we have used the fact that the incident field entering the system through the cavity loss channels is in a vacuum state, and therefore $\left \langle \delta \tilde X_{C,l}^2 \right \rangle=\left \langle \delta \tilde Y_{C, l}^2 \right \rangle=1$.

Putting this all together using Eqs.~(\ref{Vi})~and~(\ref{sumeq}), we finally arrive at an expression for the variance of the measured photocurrent
\begin{eqnarray}
\left \langle \left | \delta i \right |^2 \right \rangle &=& \frac{4 \eta^2 {g_0'}^2  \alpha_{A, \rm in}^2}{\left ( 1 + {\Delta'}^2 \right )^2} \Bigg \{ 64 \eta^2 {g_0'}^2   \alpha_{A, \rm in}^2 \frac{\Delta'^2}{\left ( 1+ \Delta'^2 \right ) ^4}  \left \langle {{g_{\rm therm}^{\rm self}}'} ^2 \right \rangle  + \frac{16 \eta^2  \alpha_{A, \rm in}^2}{\left ( 1 + {\Delta'}^2 \right )^2}   \left \langle {{R_{\rm therm}^{\rm cross}}'} ^2 \right \rangle \nonumber \\
&& + 16 {g_0'}^2 \alpha_{A, \rm in}^2 \left [ \frac{ 2 \eta - 1   -{\Delta'}^2 }{\left ( 1 + \Delta'^2 \right )^2  }\right ]^2  \left \langle { \delta \gamma_{\rm in}'  }^2 \right \rangle
+\frac{ 4  \eta^2 { g_0'}^2}{ \left (1 + \Delta'^2 \right )^2} \left [ 1 +  V_{\rm RIN} (\omega) \alpha_{A, \rm in}^2 + \frac{4 \omega'^2 \Delta'^2 }{  \left (1 + \Delta'^2 \right )^4 } V_{\zeta} (\omega) \alpha_{A, \rm in}^2 \right ]
\nonumber\\
&& + \frac{  4 \eta(1-\eta) { g_0'}^2 }{  \left (1 + \Delta'^2 \right )^2  } \left [ 1 +   \frac{4  \omega'^2 \Delta'^2 }{  \left (1 + \Delta'^2 \right )^4 } \right ] +  \frac{(1- 2 \eta)^2 + {\Delta'}^2}{1 + {\Delta'}^2} + \frac{4 \eta (1-\eta)}{1 + {\Delta'}^2}
 \Bigg \}\\
 &\approx& \left ( \frac{2 \eta {g_0'}  \alpha_{A, \rm in}}{ 1 + {\Delta'}^2 } \right )^2 \Bigg \{ \left ( \frac{2  \alpha_{A, \rm in}}{1+ \Delta'^2 } \right )^2 \Bigg [ \left (   \frac{ 4 \eta {g_0'}  \Delta'}{1+ \Delta'^2 } \right )^2  \left \langle {{g_{\rm therm}^{\rm self}}'} ^2 \right \rangle  + 4 \eta^2    \left \langle {{R_{\rm therm}^{\rm cross}}'} ^2 \right \rangle \nonumber \\
&& + 4 {g_0'}^2 \left [ \frac{ 2 \eta - 1   -{\Delta'}^2 }{1 + \Delta'^2   }\right ]^2  \left \langle { \delta \gamma_{\rm in}'  }^2 \right \rangle
+  \eta^2 { g_0'}^2 \left [   V_{\rm RIN} (\omega)  + \frac{4 \omega'^2 \Delta'^2 }{  \left (1 + \Delta'^2 \right )^4 } V_{\zeta} (\omega) \right ]  \Bigg ]+ 1 \Bigg \}  \\
 &\approx& {4 \eta^2 {g_0'}^2  \alpha_{A, \rm in}^2}  \Bigg \{ 4  \alpha_{A, \rm in}^2 \Bigg [ 16 \eta^2 {g_0'}^2  \Delta'^2  \left \langle {{g_{\rm therm}^{\rm self}}'} ^2 \right \rangle  + 4 \eta^2    \left \langle {{R_{\rm therm}^{\rm cross}}'} ^2 \right \rangle \nonumber \\
&& + 4 {g_0'}^2 \left ( { 2 \eta - 1  }\right )^2  \left \langle { \delta \gamma_{\rm in}'  }^2 \right \rangle
+  \eta^2 { g_0'}^2 \left (   V_{\rm RIN} (\omega)  + {4 \omega'^2 \Delta'^2 } V_{\zeta} (\omega) \right )  \Bigg ]+ 1 \Bigg \}  
\end{eqnarray}

The uncertainty in our estimate of $g_{\rm sig}$ is then {\it finally} given, using Eq.~(\ref{Vgest}), by
\begin{eqnarray}
\left \langle  \left | \delta {g_{\rm sig}^{\rm est}}' \right |^2 \right \rangle_{\rm back-scatter} &=& \left (  \frac{1}{64  {g_{0}'}^2 \eta^4 \alpha_{A, \rm in}^4 \cos^2 2 m ( \phi_{\rm sig} - \phi_0)}  \right ) \left ( \frac{2 \eta {g_0'}  \alpha_{A, \rm in}}{ 1 + {\Delta'}^2 } \right )^2 \Bigg \{ \left ( \frac{2  \alpha_{A, \rm in}}{1+ \Delta'^2 } \right )^2 \Bigg [ \left (   \frac{ 4 \eta {g_0'}  \Delta'}{1+ \Delta'^2 } \right )^2  \left \langle {{g_{\rm therm}^{\rm self}}'} ^2 \right \rangle  + 4 \eta^2    \left \langle {{R_{\rm therm}^{\rm cross}}'} ^2 \right \rangle \nonumber \\
&& + 4 {g_0'}^2 \left [ \frac{ 2 \eta - 1   -{\Delta'}^2 }{1 + \Delta'^2   }\right ]^2  \left \langle { \delta \gamma_{\rm in}'  }^2 \right \rangle
+  \eta^2 { g_0'}^2 \left [   V_{\rm RIN} (\omega)  + \frac{4 \omega'^2 \Delta'^2 }{  \left (1 + \Delta'^2 \right )^4 } V_{\zeta} (\omega) \right ]  \Bigg ]+ 1 \Bigg \}  \\
&=& \left (  \frac{1}{4  \eta \alpha_{A, \rm in} \cos 2 m ( \phi_{\rm sig} - \phi_0) \left ( 1 + {\Delta'}^2 \right ) }  \right )^2 \Bigg \{ \left ( \frac{2  \alpha_{A, \rm in}}{1+ \Delta'^2 } \right )^2 \Bigg [ \left (   \frac{ 4 \eta {g_0'}  \Delta'}{1+ \Delta'^2 } \right )^2  \left \langle {{g_{\rm therm}^{\rm self}}'} ^2 \right \rangle  + 4 \eta^2    \left \langle {{R_{\rm therm}^{\rm cross}}'} ^2 \right \rangle \nonumber \\
&& + 4 {g_0'}^2 \left [ \frac{ 2 \eta - 1   -{\Delta'}^2 }{1 + \Delta'^2   }\right ]^2  \left \langle { \delta \gamma_{\rm in}'  }^2 \right \rangle
+  \eta^2 { g_0'}^2 \left [   V_{\rm RIN} (\omega)  + \frac{4 \omega'^2 \Delta'^2 }{  \left (1 + \Delta'^2 \right )^4 } V_{\zeta} (\omega) \right ]  \Bigg ]+ 1 \Bigg \}  \\
&\approx& \left (  \frac{1}{4  \eta \alpha_{A, \rm in} \cos 2 m ( \phi_{\rm sig} - \phi_0)  }  \right )^2 \Bigg \{ 4  \alpha_{A, \rm in}^2 \Bigg [ 16 \eta^2 {g_0'}^2 \Delta'^2  \left \langle {{g_{\rm therm}^{\rm self}}'} ^2 \right \rangle  + 4 \eta^2    \left \langle {{R_{\rm therm}^{\rm cross}}'} ^2 \right \rangle \nonumber \\
&& + 4 {g_0'}^2 \left ( { 2 \eta - 1   -{\Delta'}^2 }\right )^2  \left \langle { \delta \gamma_{\rm in}'  }^2 \right \rangle
+  \eta^2 { g_0'}^2 \left (  V_{\rm RIN} (\omega)  + 4 \omega'^2 \Delta'^2  V_{\zeta} (\omega) \right )  \Bigg ]+ 1 \Bigg \} 
\end{eqnarray}

The optimum sensitivity is achieved if  the molecule or nanoparticle binds at a point on the circumference of the WGM resonator where it's scattered field is exactly in, or out-of, phase with that of the intrinsic scatterer, i.e. $\phi_{\rm sig} = \phi_0 + j \pi/2$ where $j$ is an integer. We then find
\begin{eqnarray}
\left \langle  \left | \delta {g_{\rm sig}^{\rm est}}' \right |^2 \right \rangle_{\rm back-scatter} &=&\frac{ 4 {g_0'}^2  \Delta'^2}{\left ( 1+ \Delta'^2 \right )^6}     \left \langle {{g_{\rm therm}^{\rm self}}'} ^2 \right \rangle  +  \frac{1}{\left ( 1+ \Delta'^2 \right )^4}    \left \langle {{R_{\rm therm}^{\rm cross}}'} ^2 \right \rangle 
 +  {g_0'}^2 \left [ \frac{ 2 \eta - 1   -{\Delta'}^2 }{\eta \left ( 1 + \Delta'^2 \right )^3  }\right ]^2  \left \langle { \delta \gamma_{\rm in}'  }^2 \right \rangle \nonumber \\
 && +   \frac{ g_0'^2}{ 4 \left ( {1+ \Delta'^2 } \right )^4}   V_{\rm RIN} (\omega)  +    \frac{{ g_0'}^2 \omega'^2 \Delta'^2 }{  \left (1 + \Delta'^2 \right )^8 } V_{\zeta} (\omega) +   \frac{1}{16  \eta^2 \alpha_{A, \rm in}^2 \left ( 1 + {\Delta'}^2 \right )^2 }   \\
&\approx&   \frac{1}{2 \left ( 1+ \Delta'^2 \right )^4}      \left \langle {{g_{\rm therm}^{\rm self}}'} ^2 \right \rangle 
 + {g_0'}^2 \left [ \frac{ 2 \eta - 1   -{\Delta'}^2 }{\eta \left ( 1 + \Delta'^2 \right )^3  }\right ]^2  \left \langle { \delta \gamma_{\rm in}'  }^2 \right \rangle
 +   \frac{ g_0'^2}{4 \left ( {1+ \Delta'^2 } \right )^4}   V_{\rm RIN} (\omega)  +    \frac{{ g_0'}^2 \omega'^2 \Delta'^2 }{  \left (1 + \Delta'^2 \right )^8 } V_{\zeta} (\omega)  \nonumber \\
 &&
 +   \frac{1}{16  \eta^2 \alpha_{A, \rm in}^2 \left ( 1 + {\Delta'}^2 \right )^2 }   \label{gdelta}   \\
&\approx&  
    \left (\frac{ g_0'}{2}\right )^2  V_{\rm RIN} (\omega)  +    { { g_0'}^2 \omega'^2 \Delta'^2 } V_{\zeta} (\omega) + {g_0'}^2 \left ( \frac{ 2 \eta - 1  }{\eta}\right )^2  \left \langle { \delta \gamma_{\rm in}'  }^2 \right \rangle+ \frac{1}{2} \left \langle {{g_{\rm therm}^{\rm self}}'} ^2 \right \rangle +   \frac{1}{16  \eta^2 \alpha_{A, \rm in}^2  }   
\end{eqnarray}
where we have used Eq.~(\ref{therm_relation}) to relate the self- and cross- thermorefractive noise. 
Comparing this to the quantum noise limit of standard single optical mode dispersive sensing given in Eq.~(\ref{QNLsense}) we see some striking similarities and differences. Firstly, let us examine the fundamental noise sources, shot noise and thermorefractive noise, given respectively in the last, and second last, terms in both expressions. We observe that {\it exactly} the same shot noise limit is reached by directly measuring back-scatter, as is achieved with a more complicated phase measurement on the single optical mode case. Furthermore, we observe that the thermorefractive noise is reduced by a factor of two.  This can be understood since when thermorefractive fluctuations scatter light, only the components of the scattered field in, or out-of, phase with the intrinsic scatterer are observed. Thus, only thermorefractive noise from a smaller sample of the bulk material of the WGM resonator is sampled. The third term in both expressions results from input coupling noise, which may also be fundamental if the noise arises from thermal fluctuations in the coupling rate. Here, remembering that $\Delta' = \Delta_0' + g_0'$, we can see that again the noise contributions using one, or two, optical modes, are very similar. Here, however, detection with a single mode has an advantage, since $\Delta_0'$ may be set such that $\Delta'=0$, in principle fully suppressing the input coupling fluctuations. In the backscatter (two mode) case, this is not possible.  However, the input coupling contribution may still be perfectly suppressed by sitting exactly on critical coupling $\eta=0.5$. The second term in both expressions is due to the frequency (or phase) noise on the light.  In the one mode case, it appears that the phase noise can be removed by sitting at critical coupling, however, the reason for this is that, at critical coupling, no light (on average) exits the resonator. An optical local oscillator is then required both to boost the signal up to measurable levels and to provide a phase reference. Phase noise in this local oscillator will also introduce phase noise in the measurement. In the two mode case, direct detection is used, without the requirement of a local oscillator. Furthermore, by setting $\Delta'=0$ the phase noise on the incident laser can be entirely suppressed. The frequency noise suppression achieved with back-scatter sensing is shown as a function of detuning $\Delta'$ in the inset of Fig.~2 of the main paper, with a fit to the detuning dependence of the co-efficient in front of $V_{\zeta}$ in Eq.~(\ref{gdelta}) yielding excellent agreement. Finally, the first term in both expressions is the relative intensity noise.  Here the situation is reversed, relative intensity noise can be perfectly suppressed in the one mode measurement, whilst it cannot be in the backscatter measurements. 

Converting to terminology consistent with the main paper, we have $\left \langle  \left | \delta {g_{\rm sig}^{\rm est}}' \right |^2 \right \rangle_{\rm back-scatter} = S(\omega)$, $V_{\rm RIN} = S_{\rm RIN}$, $\omega^2 V_{\zeta}=S_\omega$, $V_\gamma'=S_\gamma$, $V_{\rm therm}' =S_{\rm T} /\bar \gamma$,
 and $S_{
\rm shot} = {\bar \gamma^2}/{16 \eta^2   n_{\rm in}}$ so that
\begin{equation}
S(\omega) =\left ( \frac{g_0}{2} \right )^2 S_{\rm RIN} (\omega)  +  \left ( \frac{g_0 \Delta}{\bar \gamma}\right )^2 S_{\omega} (\omega)  +    g_0^2\left ( \frac{1 -2 \eta }{\eta}  \right )^2  {S_\gamma(\omega)}  
 +  \frac{1}{2} S_{\rm T} (\omega) + S_{\rm shot},
\end{equation}
where $S_\omega (\omega)$ is the laser frequency noise with it's relation to laser phase noise given, for example, in Ref.~\cite{three}. Furthermore, in the experiments reported in the paper the optical coupling is set to critical coupling such that $\eta=0.5$, consequently the contribution of input coupling fluctuations is eliminated. Taking this case, and setting $\gamma=\bar \gamma$ we have
\begin{equation}
S(\omega) =\left ( \frac{g_0}{2} \right )^2 S_{\rm RIN} (\omega)  +  \left ( \frac{g_0 \Delta}{\gamma}\right )^2 S_{\omega} (\omega) 
 +  \frac{1}{2} S_{\rm T} (\omega) + S_{\rm shot},
\end{equation}
which is Eq.~(1) of the paper.

\end{document}